\documentclass[10pt,lettersize,journal]{IEEEtran}

\usepackage{amsmath,amsfonts}
\usepackage{algorithmic}
\usepackage{algorithm}
\usepackage{balance}
\usepackage{array}
\usepackage[caption=false,font=normalsize,labelfont=sf,textfont=sf]{subfig}
\usepackage{textcomp}
\usepackage{stfloats}
\usepackage{verbatim}
\usepackage{graphicx}
\usepackage{adjustbox}
\usepackage{booktabs}
\usepackage{cite}
\usepackage{pifont} 
\usepackage{xcolor} 
\newcommand{\hb}[1]{\textcolor{blue}{#1}} 
\usepackage{multirow}
\usepackage{subcaption}
\usepackage{float}
\usepackage{placeins}
\usepackage{wrapfig}
\usepackage[hidelinks,colorlinks=true,linkcolor=black,citecolor=black,urlcolor=black]{hyperref}
\usepackage{makecell}
\usepackage{tikz}
\usepackage{enumitem}
\usepackage[table]{xcolor}
\usepackage{caption}

\usepackage{arydshln}
\newif\ifcleanmode
\cleanmodetrue  

\ifcleanmode
  \definecolor{red}{named}{black}  
  \definecolor{blue}{named}{black} 
\fi

\begin{document}

\title{Detecting Masquerade Attacks in Controller Area Networks Using Graph Machine Learning}

\author{
William Marfo\textsuperscript{\textdagger}, \textit{Member, IEEE},
Pablo Moriano\textsuperscript{*}, \textit{Senior Member, IEEE},
Deepak K. Tosh\textsuperscript{\textdagger}, \textit{Senior Member, IEEE},
Shirley V. Moore\textsuperscript{\textdagger}, \textit{Member, IEEE}\\
\textsuperscript{\textdagger}Department of Computer Science, \textsuperscript{*}Computer Science and Mathematics Division\\
\textsuperscript{\textdagger}University of Texas at El Paso, El Paso, TX 79968, USA\\
\textsuperscript{*}Oak Ridge National Laboratory, Oak Ridge, TN 37830, USA\\
wmarfo@miners.utep.edu, moriano@ornl.gov, \{dktosh, svmoore\}@utep.edu
}



\maketitle

\begin{abstract}
Modern vehicles rely on a myriad of electronic control units (ECUs) interconnected via controller area networks (CANs) for critical operations. Despite their ubiquitous use and reliability, CANs are susceptible to sophisticated cyberattacks, particularly masquerade attacks, which inject false data that mimic legitimate messages at the expected frequency. These attacks pose severe risks such as unintended acceleration, brake deactivation, and rogue steering. Traditional intrusion detection systems (IDS) often struggle to detect these subtle intrusions due to their seamless integration into normal traffic. This paper introduces a novel framework for detecting masquerade attacks in the CAN bus using graph machine learning (ML). We hypothesize that the integration of shallow graph embeddings with time series features derived from CAN frames enhances the detection of masquerade attacks. We show that by representing CAN bus frames as message sequence graphs (MSGs) and enriching each node with contextual statistical attributes from time series, we can enhance detection capabilities across various attack patterns compared to using graph-based features only. Our method ensures a comprehensive and dynamic analysis of CAN frame interactions, improving robustness and efficiency. Extensive experiments on the ROAD dataset validate the effectiveness of our approach, demonstrating statistically significant improvements in the detection rates of masquerade attacks compared to a baseline that uses graph-based features only as confirmed by Mann-Whitney \(\mathrm{U}\) and Kolmogorov-Smirnov tests $(p < 0.05)$.
\end{abstract}


\section{Introduction} \label{sec:Introduction}

Controller area networks (CANs) have become ubiquitous in modern vehicles and industrial applications, forming the backbone of communication between electronic control units (ECUs) responsible for essential functions such as acceleration, braking, and steering \cite{halder2020coids, Jedh2021, shahriar_cantropy}. Renowned for their reliability and efficiency, CANs have become the de facto standard for in-vehicle communication. However, increased integration with external systems—such as diagnostics, firmware updates, and advanced driver assistance systems (ADAS)—has exposed CANs to a growing range of cyber threats that compromise data integrity and system functionality \cite{Moriano:2022:Signal:Based:IDS}. CAN attacks are typically grouped into fabrication, suspension, and masquerade categories~\cite{verma2022addressing}.

Among these threats, masquerade attacks are particularly insidious. These attacks inject malicious CAN frames that emulate legitimate messages, manipulating vehicle behavior while evading conventional intrusion detection systems (IDS) \cite{Miller2015, sharmin2023comparative}. Unlike fabrication or suspension attacks, which create noticeable disruptions, masquerade attacks mirror the structure and frequency of authentic messages, making them exceptionally difficult to detect \cite{verma2022addressing}. The consequences are severe: a compromised ECU could subtly trigger unintended acceleration, disengage braking mechanisms, or disable safety-critical features—actions that, if left unnoticed, can lead to catastrophic outcomes on the road.

While IDS solutions have been deployed to safeguard CANs, most fall into two categories: signature-based and anomaly-based approaches. Signature-based IDS are effective against known threats but fail against zero-day attacks \cite{marfo2022condition, wang2024intrusion}. Anomaly-based IDS offer broader coverage but struggle with high rates of both false positives and negatives, as well as insufficient context-awareness for detecting stealthy behavior \cite{sun2021anomaly, Xiao2023}. Existing solutions also tend to rely on either structural or signal-based analysis in isolation, failing to capture the critical relationship between message sequences and signal dynamics—a relationship that is essential for detecting masquerade attacks.

{\color{blue}Masquerade attacks are uniquely challenging to detect due to their ability to seamlessly blend into normal traffic. These attacks mimic the frequency and content of legitimate frames, thereby bypassing traditional IDS techniques that rely on surface-level anomalies. Prior studies have shown that detection requires examining multiple consecutive frames, as single-message checks are often insufficient \cite{Jedh2021}. Moreover, methods that use either graph structure or signal statistics alone often miss subtle patterns that span both temporal and structural domains \cite{moriano2024benchmarkingunsupervisedonlineids, hanselmann2020canet}. For example, a vehicle under attack may gradually exhibit abnormal behavior—such as slowing to a halt—not due to one erroneous message, but due to a sequence of seemingly valid ones. Detecting such behavior demands a multimodal perspective.}

To address these limitations, this work proposes a unified framework that integrates graph-based and time series analysis for the detection of masquerade attacks in CANs. By modeling CAN traffic as MSGs and enriching each node with contextual time series features, our method captures both structural irregularities and signal-level deviations. This dual perspective enables robust detection of subtle intrusions that would otherwise go unnoticed in traditional systems. We also report the testing time per window (TTW)~\cite{nichelini2023canova} as a proxy for inference time to assess the real-time detection capability of our approach.

The main contributions of this paper can be summarized as follows:
\begin{itemize}[left=0pt]
\item We propose a comprehensive framework for detecting masquerade attacks in the CAN bus. Our approach uses graph ML by integrating shallow graph embeddings and time series analysis to capture both the structural and temporal aspects of CAN frames. By representing CAN frames as MSGs, we enhance the ability to detect subtle masquerade attacks that traditional IDS might miss.

    \item We implement a robust node annotation technique within MSGs that significantly enhances the detection process. By incorporating key statistical attributes derived from time series data, such as the mean and standard deviation of signals, into each CAN ID node, our method provides a richer, context-aware analysis. This detailed annotation not only improves the accuracy of masquerade attack detection but also ensures continuous monitoring and dynamic adaptation to changing network behaviors.
    
    \item We evaluate our proposed framework through extensive experiments using the ROAD dataset, a benchmark in vehicular network security research. Our results demonstrate significant improvements in detecting masquerade attacks when compared to an approach based graph topology only. The evaluation highlights the scalability and robustness of our approach, ensuring its practical applicability in modern vehicular networks.
\end{itemize}

We have made the code available to reproduce all the results at \cite{GraphML-CONTROLLER-AREA-NETWORK-Attack-Detection}.

\section{Background}
\label{sec:Background}

\subsection{CAN Protocol}

\begin{figure}[H]
\centering
\includegraphics[width=0.70\linewidth, trim=10 10 10 10, clip]{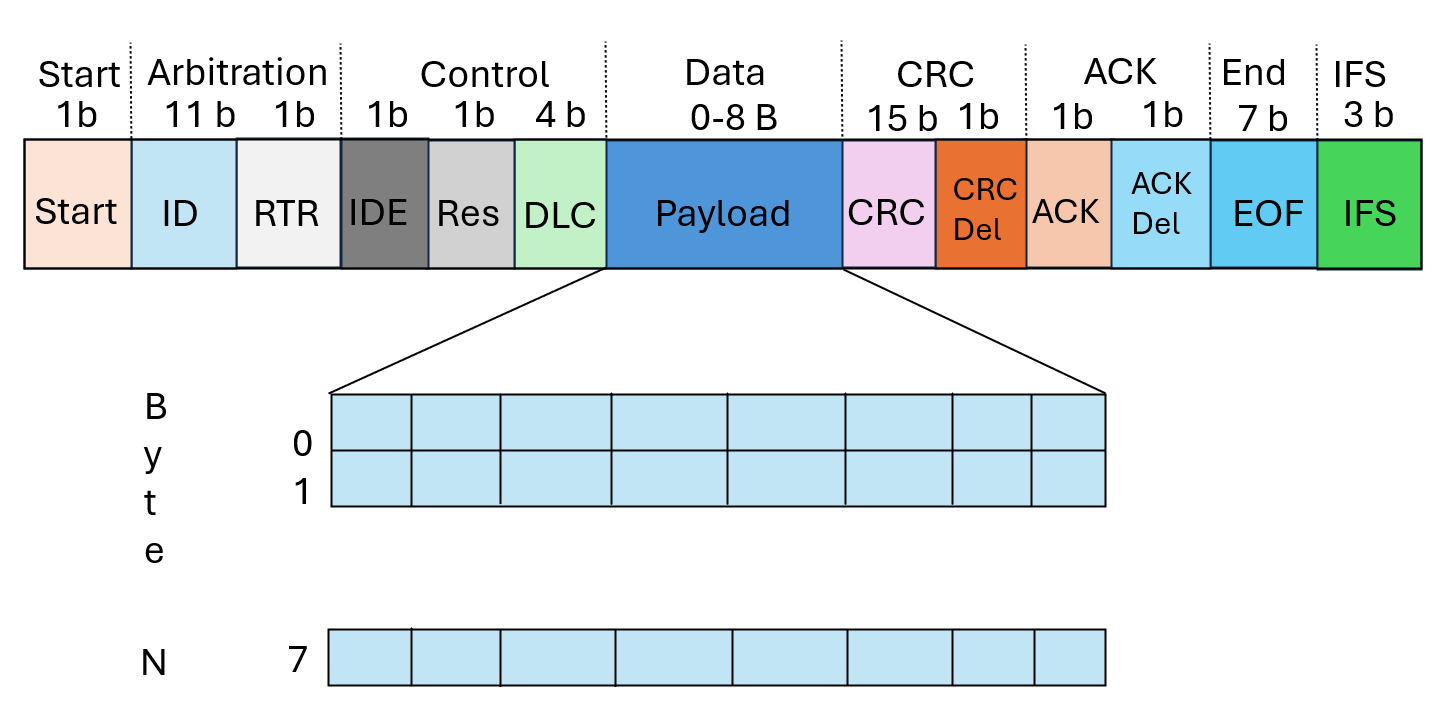}
\caption{CAN protocol frame with 8-byte payload. The full structure is presented for comprehensive understanding, with a specific focus on the ID and payload fields for our analysis.}
\label{fig:can_frame}
\end{figure}

\color{blue} CAN is a widely adopted message-based communication protocol designed to enable efficient data exchange among ECUs in vehicles and industrial systems~\cite{DiNataleZengGiustoGhosal2012CAN}. CAN operates on the physical and data-link layers of the OSI model, using a multi-master, broadcast-based communication model over a two-wire bus, as shown in Fig.~\ref{fig:can_frame}.

Each CAN frame contains multiple fields; however, in the context of this work, we focus on two: the 11-bit arbitration ID and the up-to-8-byte data field. The ID determines message priority, with lower values corresponding to higher priority. The data field holds signal-level information such as speed or engine temperature. Messages with the same ID are transmitted periodically with a fixed payload structure, enabling the modeling of temporal and structural relationships among signals. For example, a powertrain ECU may broadcast a frame with ID 0x102 at 20 Hz containing real-time values for speed and odometer. These periodic patterns are typically defined using a CAN database (DBC) file, which decodes binary payloads into meaningful physical signals. This representation is particularly useful for downstream ML-based IDS that leverage time series analysis and graph-based representations~\cite{Zhang2023}.

Beyond the protocol's operational details, it is crucial to understand the security landscape and the various types of attacks targeting the CAN bus. Our approach targets multiple tiers of CAN bus threats, which are typically classified into three categories based on adversarial intent~\cite{verma2022addressing}:
\begin{itemize}[left=0pt]
    \item \textbf{Fabrication attacks:} Malicious messages are injected onto the bus by a rogue ECU, without disrupting normal traffic from other ECUs~\cite{gazdag2023crysys}.
    \item \textbf{Suspension attacks:} A legitimate ECU is disabled or disconnected, causing its expected messages to vanish temporarily~\cite{rajapaksha2024canmirgu}.
    \item \textbf{Masquerade attacks:} The most stealthy and severe—these suppress a legitimate ECU and impersonate it by replaying spoofed frames at realistic timing and content patterns~\cite{Moriano:2022:Signal:Based:IDS}.
\end{itemize}

Note that while this study focuses on masquerade attacks due to their stealth and impact, the proposed framework is also applicable to fabrication and suspension attacks, as these introduce structural and temporal deviations captured by MSGs.

\color{black}
\subsection{CAN Message Sequence Graph (MSG)}

The MSG is a crucial tool for analyzing the behavior of the CAN \cite{Jedh2021}. It captures the sequence pattern of messages, reflecting their normal sequence of messages in CAN. Nodes in this graph correspond to unique CAN identifiers (IDs), denoting different ECUs or functions within the vehicle network. Edges map the sequence of messages between these IDs illustrating how information flows within the system. The MSG models regular patterns of communication between ECUs within specified sliding windows, allowing for detailed observation of both normal and potentially malicious activities. Typical sequence patterns found during various vehicle operations, such as parking or acceleration, can be captured by the MSG. These patterns form the basis for detecting deviations that may indicate cybersecurity threats like fabrication and suspension attacks.

Constructing an MSG involves parsing time-stamped CAN messages into a graph structure that captures the sequence and frequency of CAN ID transitions. This highlights communication patterns typical of normal vehicle operations—such as fuel delivery and speed coordination during acceleration—and enables detection of disruptions indicative of attacks. {\color{blue}As shown in Section~IV-A, MSGs visualize these dynamics, supporting robust analysis of ECU interactions and vehicular communication integrity.}

\section{Related Work} \label{sec:relatedwork}

\newcommand{\cmark}{\textcolor{green}{\ding{51}}} 
\newcommand{\xmark}{\textcolor{red}{\ding{55}}}   

\begin{table*}[t]
\centering
\caption{COMPARISON OF RELATED WORKS ON INTRUSION DETECTION IN CANS.}
\label{tab:relatedwork}
\begin{adjustbox}{max width=\textwidth}
\begin{tabular}{@{}lccccccc@{}}
\toprule
\textbf{Reference} & \textbf{Graph-based} & \textbf{Graph embeddings} & \textbf{Time series features} & \textbf{Node annotation} & \textbf{Dataset} & \textbf{Evaluation metrics} & \textbf{Public code/Data} \\
\midrule

Islam et al. [2020] \cite{Islam2020} & \cmark & \xmark & \xmark & \xmark & HCRL & Misclassification rate & \cmark \\
\hdashline
Hanselmann et al. [2020] \cite{hanselmann2020canet} & \xmark & \xmark & \cmark & \xmark & Custom & AUC-ROC & \xmark \\
\hdashline
Jedh et al. [2021] \cite{Jedh2021} & \cmark & \xmark & \xmark & \xmark & Custom & Accuracy, Detection time & \xmark \\
\hdashline
Islam et al. [2022] \cite{Islam2022} & \cmark & \xmark & \xmark & \xmark & OpelAstra & Accuracy, Training time & \xmark \\
\hdashline
Sreelekshmi et al. [2022] \cite{sreelekshmi2022graph} & \cmark & \xmark & \xmark & \xmark & Car Hacking 2020 & Accuracy & \cmark \\
\hdashline
Refat et al. [2022] \cite{Refat2022} & \cmark & \xmark & \xmark & \xmark & Real CAN Data & Accuracy & \xmark \\
\hdashline
Shahriar et al. [2022] \cite{shahriar2022canshield} & \xmark & \xmark & \cmark & \xmark & ROAD, SynCAN & AUC-ROC & \cmark \\
\hdashline
Zhang et al. [2023] \cite{Zhang2023} & \cmark & \cmark (GNN) & \xmark & \cmark & Ford Transit 500 & Accuracy & \xmark \\
\hdashline
Park et al. [2023] \cite{Park2023} & \cmark & \xmark & \xmark & \xmark & Car Hacking 2020 & Accuracy & \xmark \\
\hdashline
Xiao et al. [2023] \cite{Xiao2023} & \cmark & \cmark (GAT) & \xmark & \xmark & HCRL & Accuracy & \xmark \\
\hdashline
Meng et al. [2023] \cite{Meng2023} & \cmark & \xmark & \xmark & \xmark & OTIDS & Accuracy & \xmark \\
\hdashline
Moriano et al. [2024] \cite{moriano2024benchmarkingunsupervisedonlineids} & \xmark & \xmark & \cmark & \xmark & ROAD & AUC-ROC, TTW & \cmark \\
\hdashline
\rowcolor{gray!15}
\textbf{Our Work} & \cmark & \cmark (node2Vec) & \cmark & \cmark & ROAD & AUC-ROC, FPR, FNR, TTW & \cmark \\

\bottomrule
\end{tabular}
\end{adjustbox}
\end{table*}

We discuss prior work closely related to graph-based CAN IDS (Section III-A) and other prior work related to general CAN IDS for detecting masquerade attacks (Section III-B).

\subsection{Prior Work Closely Related to the Present Study}

Islam et al. \cite{Islam2020} proposed a four-stage IDS utilizing the chi-squared method to identify both strong and weak cyberattacks on the CAN bus. Their approach, which is among the first graph-based defense systems for CAN, showed misclassification rates of 5.26\% for DoS attacks, 10\% for fuzzy attacks, 4.76\% for replay attacks, and zero misclassification for spoofing attacks. The dataset used for evaluation was obtained from the Hacking and Countermeasure Research Lab \cite{Lee2018CANDataset}. This study emphasized the need for robust security mechanisms in modern vehicles and demonstrated superior accuracy compared to existing ID sequence-based methods.

Jedh et al. \cite{Jedh2021} proposed a message injection attack detection solution. Their approach leveraged MSGs and used graph-based analytics and anomaly detection to detect malicious message injections with high accuracy and low detection time. They validated their approach using a dataset collected from a moving vehicle under attack conditions. This study highlighted the importance of addressing the security of in-vehicle communication networks without relying on proprietary ECU information.



Refat et al. \cite{Refat2022} presented a lightweight IDS that translates CAN traffic into temporal graphs and applies neighborhood-based graph similarity techniques to detect message injection attacks. The system was evaluated using real vehicle data \cite{Song2020InVehicleNetwork}, achieving a detection accuracy of 96.01\% for spoofing, fuzzy, and DoS attacks. This work contributed to vehicle security by providing a computationally efficient method without requiring changes to the CAN protocol.

Zhang et al. \cite{Zhang2023} proposed a CAN bus anomaly detection system using graph neural networks (GNNs) to detect message injection, suspension, and falsification attacks. Their approach involved creating directed attributed graphs from CAN message streams and training a two-stage classifier cascade. The evaluation on a Ford Transit 500 dataset \cite{Othmane2022CANBusInjection} demonstrated the system's efficiency in real-time detection and its ability to handle new anomalies through federated learning training.


Xiao et al. \cite{Xiao2023} proposed the CAN-GAT model based on graph attention networks for in-vehicle networks. By transforming CAN bus messages into graph structures, the model captured the correlation between traffic bytes, improving detection accuracy and efficiency. The model was evaluated using datasets from the Hacking and Countermeasure Research Lab \cite{Lee2018CANDataset}, demonstrating superior performance among compared GNNs. 

Meng et al. \cite{Meng2023} developed GB-IDS, an IDS leveraging a novel graph structure and a variational autoencoder for training classifiers without negative samples. Their system, evaluated on the OTIDS dataset \cite{Lee2017OTIDS}, achieved high detection success rates for DoS, fuzzing, and impersonation attacks. This study addressed the limitations of traditional IDS, i.e., \cite{Refat2022} by avoiding the need for protocol parsing and large training datasets.

{\color{blue}Compared to prior works that focus on either graph-based or time series-based CAN IDS, our approach offers a unified framework that addresses their respective limitations. Graph-based methods, such as \cite{Islam2020, Jedh2021}, use MSGs to detect structural anomalies but lack granularity for subtle signal deviations. Time series-based techniques such as \cite{moriano2024benchmarkingunsupervisedonlineids}, \cite{hanselmann2020canet} capture temporal anomalies but often miss intricate message sequence patterns and graph topology that are crucial for uncovering subtle masquerade attacks. These single-modality designs are insufficient for detecting stealthy masquerade attacks. Our framework combines graph embeddings and time series features, enabling detection across both structural and signal domains. Additionally, its flexibility—adjustable via window length and offset—supports fine-tuning in constrained environments. We evaluate this framework on the ROAD dataset~\cite{verma2022addressing}, demonstrating improved performance over recent approaches~\cite{shahriar2022canshield, Zhang2023}.}

\subsection{Other Prior Work Related to the Present Study}

Hanselmann et al. \cite{hanselmann2020canet} proposed CANet, a neural network architecture for detecting intrusions on the CAN bus. CANet, the first deep learning-based IDS capable of handling the high dimensionality of CAN bus data, was evaluated on real and synthetic CAN data. The method outperformed previous ML-based approaches with a high true negative rate, typically over 0.99, demonstrating robustness in detecting a large number of synthetic masquerade attacks \cite{hanselmann2020canet}.

Moriano et al. \cite{Moriano:2022:Signal:Based:IDS} focused on detecting masquerade attacks on the CAN bus by analyzing time series extracted from raw CAN frames. Using hierarchical clustering, this study demonstrated that changes in cluster assignments could indicate anomalous behavior. The proposed forensic tool was tested on the ROAD dataset \cite{verma2022addressing}, showing significant differences in time series clustering similarity on benign and attack conditions.



Shahriar et al. \cite{shahriar_cantropy} proposed CANtropy, a feature engineering-based lightweight CAN IDS. CANtropy explores a comprehensive set of features from both temporal and statistical domains utilizing PCA for anomaly detection. Evaluation on the SynCAN dataset \cite{hanselmann2020canet} showed CANtropy's effectiveness, with an average AUC-ROC score of 0.992, outperforming existing DL-based baselines.  

Moriano et al. \cite{moriano2024benchmarkingunsupervisedonlineids} conducted a benchmark study on four non-deep learning unsupervised online IDS for masquerade attacks in CANs. They controlled streaming data conditions in a sliding window setting and used realistic attacks from the ROAD dataset \cite{verma2022addressing}. They noticed that there is no one-size-fits-all solution for detecting masquerade attacks in CAN in an online setting, i.e., there is no single algorithm that is optimal for detecting each attack type. Notably, a method that detects changes in the hierarchical structure of clusters of time series outperformed others in detecting various attack types at the expense of computational overhead.

\section{Proposed Data-Driven Methods} \label{sec:Methods}

\begin{figure}[htbp]
  \centering
\includegraphics[width=0.85\linewidth, trim=10 10 10 10, clip]{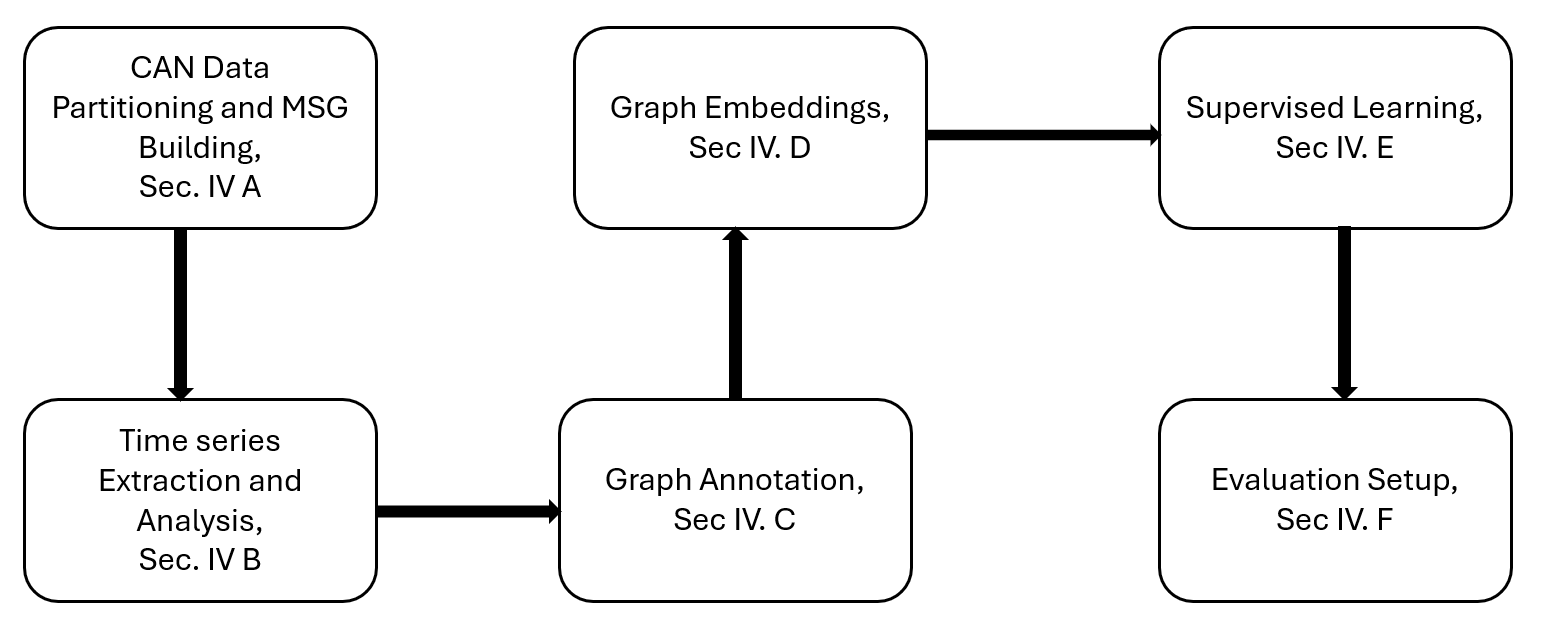}
  \caption{Proposed data-driven framework for masquerade attack detection in CAN using graph ML.}
  \label{fig:framework}
\end{figure}

This section presents our approach for detecting masquerade attacks in CAN bus communication using graph ML. We begin by outlining the complete pipeline in Fig.~\ref{fig:framework}, followed by a high-level algorithmic workflow in Algorithm~\ref{alg:master}. We then provide detailed procedures for MSG construction (Section~IV-A), time series extraction (Section~IV-B), node annotation (Section~IV-C), graph embedding generation (Section~IV-D), and supervised learning
process (Section~IV-E). Evaluation setup and metrics are described in Section~IV-F.

{\color{blue}
\textbf{Threat model:} In our framework, we assume the attacker has gained access to the CAN bus, either physically or remotely~\cite{checkoway2011comprehensive}. Masquerade attacks, as represented in the ROAD dataset~\cite{verma2022addressing}, typically involve two compromised ECUs: a \emph{weak attacker} that learns the frequency of messages sent by a targeted ECU and suppresses its transmissions, and a \emph{strong attacker} that injects spoofed messages at the expected frequency using the same ID. The coordinated nature of this attack enables the adversary to mimic legitimate traffic patterns while altering the message content. In this study, our detection framework is evaluated under the assumption that the adversary can play both attacker roles, consistent with the ROAD dataset's multi-stage masquerade attack model.}

\begin{algorithm}[h]
\caption{\color{red}End-to-end masquerade attack detection}
\label{alg:master}
\color{red}\begin{algorithmic}[1]
\REQUIRE Raw CAN logs, $\omega_t$, $\delta_t$, DBC mappings, $node2vec_{params}$
\ENSURE Trained detection model
\STATE $\mathcal{T} \gets \text{ExtractTimeSeries}(CAN, DBC)$ \hfill $\triangleright$ Sec.~IV.B
\STATE $\mathcal{G} \gets \text{Build MSGs}(CAN, \omega_t, \delta_t)$ \hfill $\triangleright$ Alg.~\ref{alg:msg_creation}
\FOR{each graph $G_i \in \mathcal{G}$}
    \STATE $\text{AnnotateNodes}(G_i, \mathcal{T})$ \hfill $\triangleright$ Alg.~\ref{alg:node_annotation}
    \STATE $E_i \gets \text{node2vec}(G_i, node2vec_{params})$ \hfill $\triangleright$ Sec.~IV.D
    \STATE $F_i \gets \text{Combine}(E_i, G_i.\mu, G_i.\sigma)$ \hfill $\triangleright$ Feature fusion
\ENDFOR
\STATE $\mathcal{M} \gets \text{TrainRandomForest}(\{F_1,...,F_n\})$
\RETURN $\mathcal{M}$
\end{algorithmic}
\end{algorithm}

{\color{blue}Note that to quantify the computational footprint of our end-to-end detection pipeline, we analyze its complexity.}

{\color{red}\noindent\textbf{\\ Computational Complexity:} The total cost is dominated by:
\begin{align*}
\mathcal{T}_{total} &= \underbrace{O\left(\frac{n}{\delta_t}\cdot\omega_t\right)}_{\text{MSG building}} + \underbrace{O(n + k \cdot s_{sig})}_{\text{Time series processing}} \\ 
&\quad + \underbrace{O(r \cdot l \cdot k)}_{\text{Graph embeddings}} + \underbrace{O(t \cdot d \cdot n_{train})}_{\text{ML Training}}
\end{align*}

\setlist[itemize]{left=0.5em, label=•}

\noindent Where:
{\color{blue}\begin{itemize}[left=1em, label=•]
    \item $n$: Total CAN messages.
    \item $k$: Unique CAN IDs.
    \item $s_{\text{sig}}$: Maximum number of signals per CAN ID (not constant; varies across IDs, typically 2–10, with an upper limit of 16 in the ROAD dataset).
    \item $n_{\text{train}}$: Number of training samples (approximately the number of sliding windows).
    \item $\omega_t$, $\delta_t$: Window size and offset for time-based configurations (seconds); $\omega_t \in [2, 15]$, $\delta_t \in [1, \omega_t]$.
    \item $\omega_s$, $\delta_s$: Window size and offset for sample-based configurations (messages); $\omega_s \in \{50, 100, \ldots, 400\}$, $\delta_s \in \{50, 100, \ldots, \omega_s\}$.
    \item $r$: Number of node2vec walks per node (100).
    \item $l$: Length of each node2vec walk (15).
    \item $t$: Number of trees in the Random Forest classifier (100).
    \item $d$: Maximum tree depth in the Random Forest classifier ($\approx \log_2 n_{\text{train}}$)".
\end{itemize}}

\noindent\textbf{Key Observations:}
\begin{itemize}
    \item Training cost is negligible because:
    \begin{itemize}
        \item It is done offline (not during vehicle operation)
        \item Typically $n_{\text{train}} \ll 10^4$ windows
        \item Random Forest scales linearly with $n_{\text{train}}$
    \end{itemize}
    \item Time series extraction is linear in $n$ (CAN-D decoding)
    \item MSG construction dominates when $\omega_t \gg \delta_t$
    \item Embedding generation (approximately 70\% runtime) scales with $k^2$. 
\end{itemize}  
}

\subsection{CAN Data Partitioning and MSG Building}
\begin{figure}[h]
    \centering
    \includegraphics[width=1\linewidth, trim=0 60 0 60, clip]{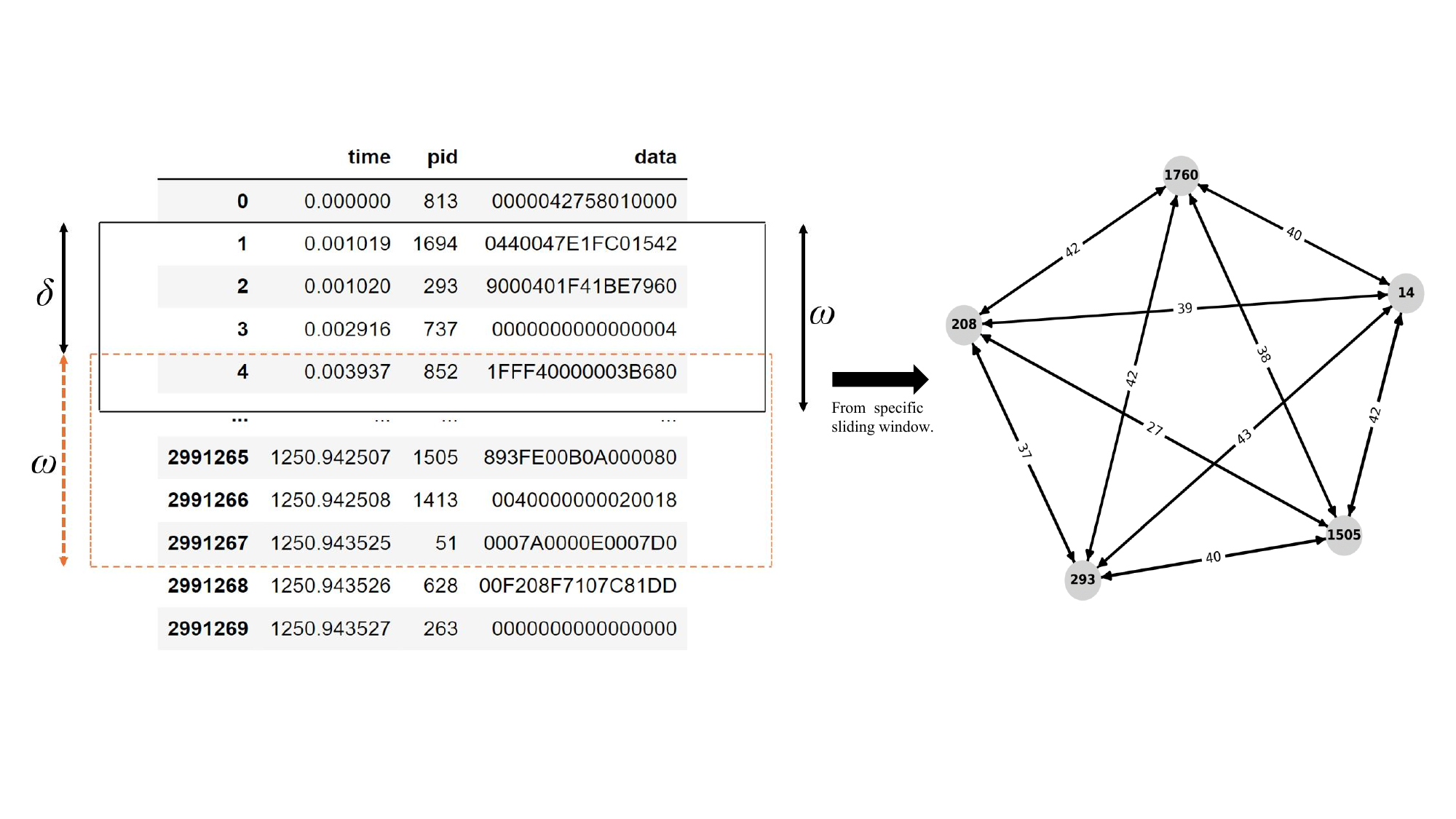}
    \caption{Sliding window partitioning of CAN messages (left) and resulting MSG subgraph (right). Top five nodes by connectivity are highlighted; edges are weighted.}
    \label{fig:can_data_to_graph_viz}
\end{figure}

We use sliding windows of fixed size to partition and process the stream of CAN data \cite{Qin2020}. The raw CAN data is first transformed into a structured dataframe format where each row represents a message with its timestamp, process ID, and data payload (see Fig. \ref{fig:can_data_to_graph_viz}). This data is then partitioned into discrete time slices defined by specific window sizes and offsets, enabling a continuous analysis of CAN data.
\\We now define the key parameters controlling the sliding windows:

\begin{itemize}[left=0pt]
    \item \textbf{Window size ($\omega$, either $\omega_t$ for time or $\omega_s$ for samples):} The length of each window determining the extent of data included in each graph representation.
    \item \textbf{Offset ($\delta$, either $\delta_t$ for time or $\delta_s$ for samples):} The sliding step between consecutive windows, where $\delta$ represents the number of samples separating each window, ensuring no loss of data and the capture of communication patterns that may span multiple windows.

\end{itemize}

{\color{red}
\noindent\textbf{Parameter Rationale—Aligning with Attack Dynamics:} The masquerade attacks in the ROAD dataset \cite{verma2022addressing} are sustained, stealthy intrusions (e.g., \texttt{correlated\_signal}, \texttt{max\_speedometer},
\texttt{max\_engine\_coolant\_temp}, \texttt{reverse\_light\_off}, \texttt{reverse\_light\_on}) lasting approximately 17–72 seconds. Our parameters were empirically optimized to capture attack duration while maintaining computational feasibility. For example, a 6-second window ensures sufficient temporal context to detect anomalies that unfold gradually, while a 3.2-second offset prevents missed detections near window boundaries through overlapping windows. This offset does not introduce a reaction delay but strategically balances attack coverage and computational feasibility—smaller offsets (e.g., 1s) increase processing load, whereas larger offsets (e.g., 6s) risk missing attacks. The selected window parameters ensure that our detection method aligns with real attack lifecycles.}

{\color{red}
\noindent\textbf{Motivation for Sliding Window Detection:} Masquerade attacks cannot be detected at the message level—they mimic legitimate traffic in short bursts but manifest as cumulative deviations over time. As demonstrated by Jedh et al. \cite{Jedh2021}, stealthy message injections, especially masquerade attacks, necessitate analysis over multiple consecutive frames rather than relying on single-message checks. Our approach aggregates e.g., 600 messages (at 100Hz) within a 6-second window, enabling structural (graph-based) and temporal (statistical) analysis to expose subtle attack patterns.}

A MSG is constructed within each defined sliding window to model CAN activity following the procedure introduced by \cite{Jedh2021} and \cite{Islam2020}. Graphs are empirically recognized as effective tools to model relationships among relational data that are too complex for tabular representation or other simpler data structures \cite{Bryan2023}. The process of building MSGs is crucial as it sets the stage for subsequent analyses, including node annotation and embedding generation. By modeling CAN data using MSG on sliding windows, we enhance our ability to detect and respond to potential masquerade attacks by finetuning the portion of analyzed data. We now detail each of the components of the MSG:

\begin{itemize}[left=0pt]
    \item \textbf{Nodes:} Each node in the MSG corresponds to a unique ID in CAN representing the header of the CAN frame. This alignment allows for the detailed representation of each frame's interactions on streams of CAN data.
    \item \textbf{Edges:} Directed edges are established based on the sequential relationships of messages. That is, an edge is formed when one ID follows another, indicating the flow of communication \cite{Islam2020}. This method effectively maps sequences of CAN messages enabling the detection of irregular patterns that may signify an attack. Edge weights are assigned to edges of MSGs patterns to quantify the frequency of observed communication sequences within a sliding window. Specifically, the weight of an edge from node \(v_i\) to node \(v_j\) is computed as the number of times the CAN message with ID \(v_j\) directly follows the CAN message with ID \(v_i\) within the same sliding window. Weights help to identify anomalous recurrent patterns that could indicate security threats. 
    
\end{itemize}

\begin{algorithm}
\caption{Creation of message sequence graphs (MSGs)}
\label{alg:msg_creation}
\begin{algorithmic}[1]
\REQUIRE Raw CAN data, $\omega_t$, $\delta_t$
\ENSURE MSG representing CAN activity over time
\STATE Initialize an empty directed graph $G$
\STATE Initialize $current\_time = start\_time$ of CAN data
\WHILE{$current\_time + \omega_t \leq end\_time$ of CAN data}
    \STATE Define $start\_time = current\_time$ and $end\_time = current\_time + \omega_t$
    \STATE Extract messages in $[start\_time, end\_time]$
    \STATE Initialize a new subgraph $G_{sub}$
    \FOR{each message $m$ in $[start\_time, end\_time]$}
        \STATE Extract $id$ from $m$
        \IF{$id$ not in $G_{sub}$}
            \STATE Add $id$ as a node in $G_{sub}$
        \ENDIF
        \STATE Determine next $id$ in sequence
        \IF{edge from current $id$ to next $id$ exists}
            \STATE Increment weight of edge
        \ELSE
            \STATE Add edge with weight = 1
        \ENDIF
    \ENDFOR
    \STATE Merge $G_{sub}$ into main graph $G$ respecting $\delta$
    \STATE Increment $current\_time$ by $\delta_t$
\ENDWHILE
\RETURN $G$
\end{algorithmic}
\end{algorithm}

  Fig. \ref{fig:can_data_to_graph_viz} shows how MSGs are built and their corresponding components. Our modeling framework ensures that both structural and temporal aspects of CAN communications are captured, facilitating a comprehensive analysis of the CAN dynamics. Algorithm~\ref{alg:msg_creation} describes the creation of MSGs.

{\color{red}
To ensure the correctness of Algorithm~\ref{alg:msg_creation}, we establish loop invariants for both the outer windowing loop and inner message processing loop. A loop invariant is a property that holds true before and after each iteration, ensuring algorithmic correctness. We validate these invariants through three steps:

\begin{enumerate}
    \item \textbf{Loop Invariants}:
    \begin{itemize}
        \item \textbf{Outer Loop (Lines 3-21)}: At iteration $i$, the main graph $G$ contains all message sequences from windows $\{[t_0,t_0+\omega_t],...,[t_{i-1},t_{i-1}+\omega_t]\}$ where $t_j = start\_time + j\cdot\delta_t$
        \item \textbf{Inner Loop (Lines 7-18)}: For message $m_k$ in window $[t_i,t_i+\omega_t]$, the subgraph $G_{sub}$:
        \begin{itemize}
            \item Contains all nodes $\{id_1,...,id_k\}$ from messages $\{m_1,...,m_k\}$
            \item Maintains edge weights equal to observed message sequence frequencies
        \end{itemize}
    \end{itemize}

    \item \textbf{Verification Steps}:
    \begin{itemize}
        \item \textbf{Initialization}:
        \begin{itemize}
            \item Outer: $G=\emptyset$ before first window (trivially satisfies)
            \item Inner: $G_{sub}=\emptyset$ before first message (trivially satisfies)
        \end{itemize}
        \item \textbf{Maintenance}:
        \begin{itemize}
            \item Outer: Each iteration processes one complete window and merges $G_{sub}$ into $G$
            \item Inner: Each message updates $G_{sub}$'s nodes/edges atomically
        \end{itemize}
        \item \textbf{Termination}:
        \begin{itemize}
            \item Outer: Terminates when $current\_time+\omega_t > end\_time$ with all complete windows processed
            \item Inner: Processes all messages in each window before continuing
        \end{itemize}
    \end{itemize}

    \item \textbf{Time Complexity}:
    \begin{itemize}
        \item \textbf{Window processing}:
        \begin{itemize}
            \item $\lceil\frac{n}{\delta_t}\rceil$ windows processed
            \item Each window requires $O(\omega_t)$ operations (message extraction + graph updates)
        \end{itemize}
        \item \textbf{Edge operations}:
        \begin{itemize}
            \item Average-case $O(1)$ for edge operations (\texttt{networkx} default implementation)
            \item Each message generates exactly one edge update
            \item Worst-case $O(k)$ per operation with hash collisions ($k$=nodes)
        \end{itemize}
        \item \textbf{Total complexity}:
        \begin{itemize}
            \item $O(\frac{n}{\delta_t}\cdot\omega_t)$ for general case
            \item $O(n)$ when $\omega_t=\delta_t$ (each message processed once)
        \end{itemize}
    \end{itemize}
\end{enumerate}

Our experimental validation confirms this analysis:
\begin{itemize}
    \item Time-based windows: $\omega_t \in [2,15]$s, $\delta_t \in [1,15]$s (all combinations)
    \item Sample-based windows: $\omega_s = \delta_s \in \{50k \mid k = 1,\ldots,8\}$ messages
    \item Linear scaling confirmed for $\omega/\delta \in [0.5,2.0]$
\end{itemize}
}

\subsection{Time Series Extraction and Analysis}

\begin{figure}[h]
  \centering
  \includegraphics[width=0.95\linewidth, trim=8 8 8 8, clip]{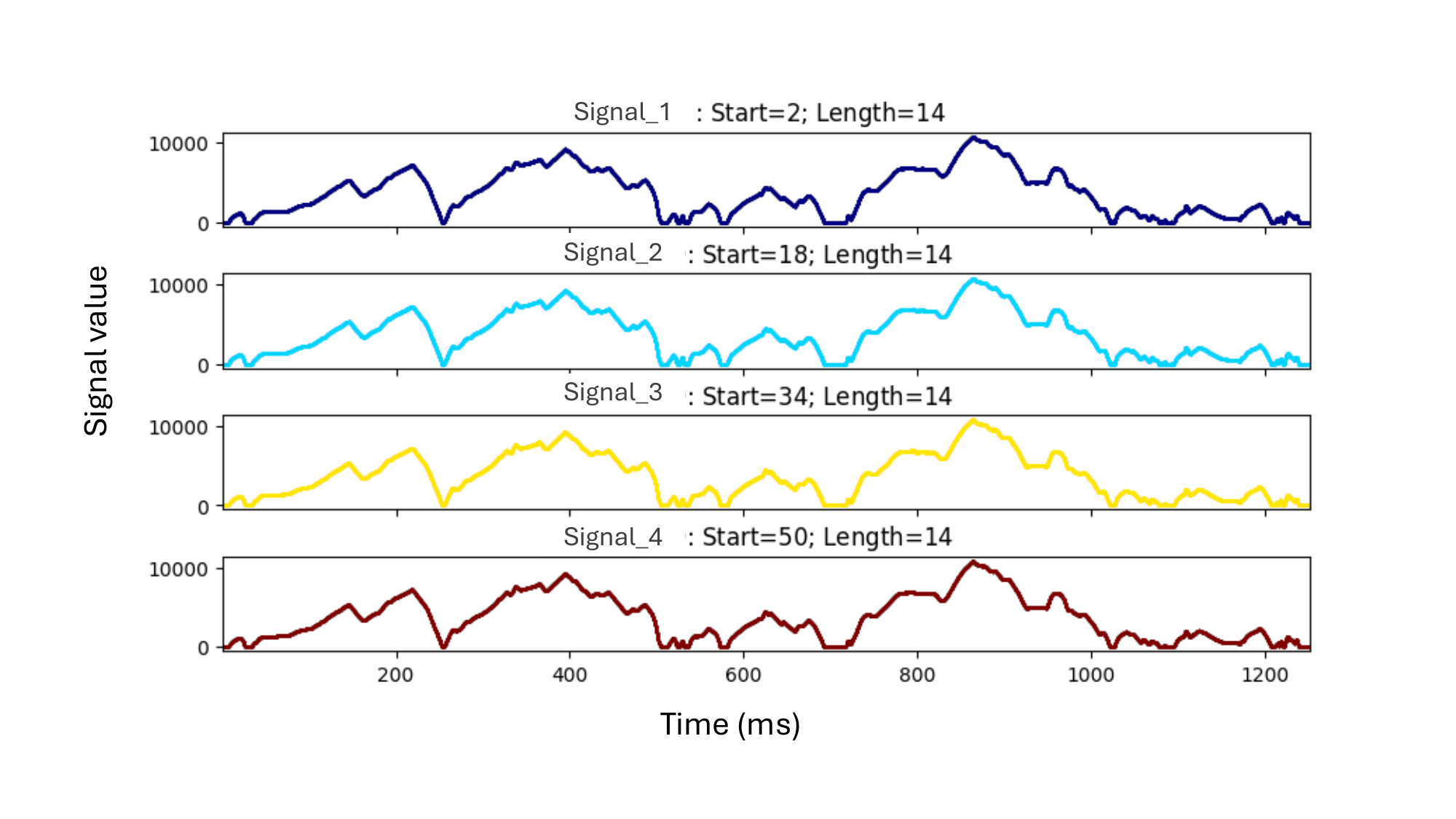}
  \caption{\color{blue}Speed signals of four wheels encoded in node ID 1760. X-axis: Time (ms); Y-axis: Signal value.}
  \label{fig:timeseries}
\end{figure}

Concurrently with the creation of the MSGs, raw CAN messages are decoded using a DBC as described in \cite{cssElectronicsDBC}. In doing so,  CAN data is transformed into a time series format to capture detailed patterns of the signals contained in every CAN ID per sliding window. We used CAN-D \cite{Verma2021} to extract timeseries from CAN logs. Up to the time of this writing, CAN-D is still the state-of-the-art method for CAN reverse engineering \cite{10092880}. Fig. \ref{fig:timeseries} shows a decoded representation of the signals representing the four wheels' speed of a vehicle \cite{verma2022addressing}. The plot is generated by translating the CAN frames into a continuous time series format, which captures the network activity over time. This representation facilitates a detailed characterization of masquerade attacks within CANs highlighting variations and trends that may not be evident relying on raw CAN data.

\subsection{Graph Annotation}

\begin{figure}[h]
  \centering
  \includegraphics[width=0.90\linewidth]{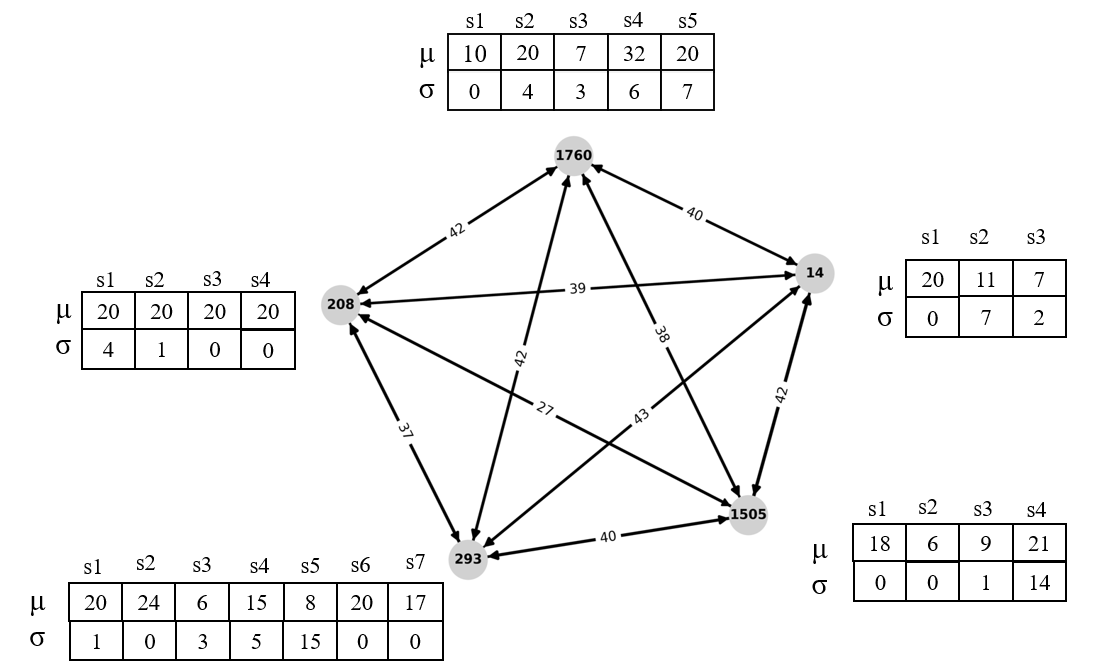}
  \caption{Illustration of graph annotation with statistical attributes in CAN. Each node represents a CAN ID and is annotated with the mean (\(\mu\)) and standard deviation (\(\sigma\)) of each signal associated with it. Note that the number of signals may vary per node.}
  \label{fig:graph_annotation}
\end{figure}

We enhance each MSG by annotating its nodes with statistical attributes derived from corresponding time series data. For each CAN ID, we compute the mean and standard deviation of its associated signals—features chosen for their simplicity and constant-time complexity, as demonstrated in prior work \cite{shahriar_cantropy}. This annotation occurs alongside MSG construction, allowing the model to jointly capture structural and temporal dynamics. Fig.~\ref{fig:graph_annotation} illustrates the annotation process, and Algorithm~\ref{alg:node_annotation} provides implementation details.

\begin{algorithm}
\caption{Node annotation with statistical attributes}
\label{alg:node_annotation}
\begin{algorithmic}[1]
\REQUIRE CAN data $\mathcal{D}$, sliding windows $\mathcal{W}$, signal mappings $\mathcal{S}$
\ENSURE Annotated graphs $\{G_t\}$ with statistical node features
\FOR{each window $w \in \mathcal{W}$}
    \FOR{each signal $s \in \mathcal{S}$}
        \STATE Extract signal values $v_s$ from $\mathcal{D}$ within window $w$
        \STATE Interpolate missing values in $v_s$ using linear interpolation
        \STATE Compute mean $\mu_s$ and standard deviation $\sigma_s$ of $v_s$
        \STATE Annotate corresponding node in $G_w$ with $(\mu_s, \sigma_s)$
    \ENDFOR
\ENDFOR
\RETURN Annotated graphs $\{G_t\}$
\end{algorithmic}
\end{algorithm}

{\color{red}\subsection{Graph Embeddings}
To model the complex interactions in CANs, we leverage graph-based techniques, specifically graph embeddings, which map network structures into low-dimensional vectors for ML.

Graph embeddings map nodes, edges, or entire graphs into a lower-dimensional vector space, preserving their structural and relational information \cite{grover2016node2vec, wu2020comprehensive}. In our study, we focus on shallow graph embeddings, specifically the \textit{node2vec} algorithm \cite{grover2016node2vec}, which is well-suited for capturing the nuances of weighted and directed graphs characteristic of MSGs. 

\textit{node2vec} generates embeddings by simulating random walks on the graph, controlled by two key parameters: the \textit{return parameter} \( p \) (likelihood of revisiting a node) and the \textit{in-out parameter} \( q \) (preference for exploring local vs. distant neighborhoods). These parameters allow the algorithm to balance the exploration of immediate connectivity and broader network contexts, making it particularly effective for modeling CAN bus data \cite{wei2024robustnessgraphembeddingmethods}. 

We configure \textit{node2vec} with 64 dimensions (\( d \)) to balance computational efficiency and feature richness \cite{Gu2021}, a walk length (\( l \)) of 15 to capture multi-hop relationships, and 100 walks per node (\( r \)) to ensure thorough exploration of the network. Additionally, we set the return parameter (\( p \)) to 1.5 and the in-out parameter (\( q \)) to 0.5, emphasizing local ECU interactions while retaining global CAN topology. This configuration ensures that the embeddings capture both the immediate communication patterns between ECUs and the broader structure of the CAN.

Unlike prior works that focus solely on node embeddings \cite{Islam2020, Jedh2021}, we construct whole-graph embeddings by averaging node embeddings within sliding windows. This aggregation process provides a unified representation of the network's behavior, capturing both structural and temporal patterns of CAN frames. Specifically, we stack the embeddings per node in the MSGs, labeling each graph to indicate normal or anomalous activity in the CAN. This approach allows us to frame our intrusion detection problem as a graph classification task \cite{Zhong2024}, combining both structural and temporal information for enhanced detection capabilities.

The relatively low embedding dimension (64) has been shown to be sufficient for capturing rich graph-related properties while maintaining computational efficiency \cite{Gu2021}, which is critical for near real-time detection of masquerade attacks in CANs. By enriching the embeddings with time-series features (e.g., mean, standard deviation) from decoded signals, we ensure a context-aware analysis of masquerade attacks, further enhancing the robustness of our framework.

The result of this process is a data frame that not only encapsulates node embeddings but also includes the augmented node attribute features with labels indicative of normal or anomalous activity states in the CAN. Before classification, we create a comprehensive feature vector for each graph by concatenating the averaged graph embeddings with the time series statistical features (mean and standard deviation of signals). This representation combines both structural (graph) and temporal (time series) information, providing a robust input for ML classifiers. These comprehensive feature vectors serve as the foundation for our supervised learning approach to masquerade attack detection.}

\subsection{Supervised Learning}
To evaluate the effectiveness of our feature representation and overall anomaly detection framework, we employ supervised learning techniques. We experiment with two ensemble learning methods: Random Forest (RF) \cite{breiman2001random} and XGBoost \cite{10.1145/2939672.2939785}. These decision tree-based ensemble learning methods are renowned for their high accuracy, scalability, and robustness, making them exceptionally suitable for dealing with the high-dimensional space characteristic of CAN data. The RF model is instantiated with 100 trees, and depth is capped at 20 to prevent overfitting. A combination of a minimum samples split of 6 and minimum samples leaf of 2 is used to control the growth of the trees further. As the training dataset exhibits a 1:10 ratio between the minority (attack) and majority (normal) classes, we balance the dataset using the synthetic minority over-sampling technique (SMOTE) \cite{10.5555/1622407.1622416}. A 5-fold stratified \emph{k}-fold cross-validation is performed to ensure the model's generalizability during training. The XGBoost model yielded similar results to the RF classifier in our experiments. Despite the comparable performance, we opted to report only results for the RF classifier due to its simpler interpretability and less intensive hyperparameter tuning process, which aligns well with the goals of this study.

\subsection{Evaluation Setup}
Evaluation is performed at the sliding window level. In doing so, we verify the proportion of sliding windows coinciding with an attack region in the testing captures, which refers to the recorded CAN bus data used for testing. We found that for the combinations of $(\omega_t, \delta_t)$ or $(\omega_s, \delta_s)$, the proportion of windows containing attacks tended to be balanced, i.e., greater than 40\%. To quantitatively assess the performance of our anomaly detection framework, we use the area under the receiver operating characteristic curve (AUC-ROC) \cite{FAWCETT2006861}. The AUC-ROC is a comprehensive evaluation metric for classification problems across various threshold settings. It reflects the model's ability to distinguish between classes, with higher values indicating better performance. The ROC curve is a plot of the true positive rate (TPR) against the false positive rate (FPR) at different threshold levels. The TPR, also known as sensitivity or recall, measures the proportion of actual positives correctly identified, while the FPR measures the proportion of actual negatives incorrectly identified as positive. Mathematically, the AUC-ROC is defined as: \(\text{AUC-ROC} = \int_{0}^{1} \text{TPR}(\text{FPR}^{-1}(x)) \, dx.\)
 
In practical terms, the AUC-ROC can be interpreted as the probability that the classifier will rank a randomly chosen positive instance higher than a randomly chosen negative instance. This interpretation is particularly useful in our context, where we aim to detect subtle masquerade attacks embedded within normal CAN traffic.

{\color{red}\textbf{Performance Metrics:} We evaluated computational performance using the testing time per window (TTW) metric by Nichelini et al.~\cite{nichelini2023canova}, which quantifies the amount of time required to process each window across attack scenarios. We define TTW as:

\[
\text{TTW} = \frac{\text{Total Detection Time}}{\text{Number of Windows}}
\]

TTW provides a practical means of evaluating the computational feasibility of our detection framework in real-time settings.} All experiments were conducted on a 12th Gen Intel Core i9-12900HK CPU with an NVIDIA RTX 3080 Ti GPU and 32GB RAM.

\subsection{Dataset}
Our framework is evaluated on the Real ORNL Automotive Dynamometer (ROAD) dataset \cite{verma2022addressing}, collected from a physical vehicle at Oak Ridge National Laboratory (ORNL). This dataset includes physically verified fabrication and simulated masquerade attacks, offering a realistic benchmark for CAN security methods. It is distinguished by its high-quality CAN data and credible attack scenarios across diverse driving conditions. The dataset comprises 3.5 hours of recorded data—3 hours for training and 30 minutes for testing. The test data incorporates five masquerade attacks including correlated signal, max engine coolant, max speedometer, reverse light off/on attacks. These attacks are designed to manipulate specific vehicle states. This dataset enables us to conduct a comprehensive evaluation of our framework under a variety of attack conditions. {\color{red}Table~\ref{tab:attack_characteristics} summarizes the durations, injection intervals, and impacts of the masquerade attacks.}

{\color{red}\begin{table}[h]
\color{red}\centering
\caption{\color{red}COMPARISON OF RELATED WORKS ON INTRUSION DETECTION IN CANS.}
\label{tab:attack_characteristics}
\begin{adjustbox}{max width=0.5\textwidth}
\begin{tabular}{@{}lccp{4.1cm}@{}}
\toprule
\textbf{Attack Name} & \textbf{Duration (s)} & \textbf{Injection Interval (s)} & \textbf{Description} \\
\midrule

\texttt{correlated\_signal\_1} & 33.10 & [9.19, 30.05] & \multirow{3}{4.1cm}{Injects varying values for wheel speeds, causing the vehicle to halt.} \\
\texttt{correlated\_signal\_2} & 28.23 & [6.83, 28.23] & \\
\texttt{correlated\_signal\_3} & 16.96 & [4.32, 16.96] & \\
\hdashline

\texttt{max\_speedometer\_1} & 88.02 & [42.01, 66.45] & \multirow{3}{4.1cm}{Injects the maximum value to be displayed on the speedometer.} \\
\texttt{max\_speedometer\_2} & 59.70 & [16.01, 47.41] & \\
\texttt{max\_speedometer\_3} & 86.77 & [9.52, 70.59] & \\
\hdashline

\texttt{max\_engine\_coolant} & 25.88 & [19.98, 24.17] & Injects the maximum value, triggering the coolant warning light. \\
\hdashline

\texttt{reverse\_light\_off\_1} & 28.11 & [16.63, 23.35] & \multirow{3}{4.1cm}{Toggles the reverse light irrespective of the actual gear position.} \\
\texttt{reverse\_light\_off\_2} & 40.67 & [13.17, 36.88] & \\
\texttt{reverse\_light\_off\_3} & 57.88 & [16.52, 40.86] & \\
\hdashline

\texttt{reverse\_light\_on\_1} & 54.85 & [18.93, 38.84] & \multirow{3}{4.1cm}{Toggles the reverse light irrespective of the actual gear position.} \\
\texttt{reverse\_light\_on\_2} & 72.02 & [20.41, 57.30] & \\
\texttt{reverse\_light\_on\_3} & 64.26 & [23.07, 46.58] & \\

\bottomrule
\end{tabular}
\end{adjustbox}
\end{table}
}

\section{Results} \label{sec:Results}

We evaluate our framework on the ROAD dataset \cite{verma2022addressing} under two settings: (1) a baseline using graph embeddings only and (2) an enhanced configuration combining graph embeddings with time series features. All methods were implemented in \texttt{Python 3.8.18}, using \texttt{NetworkX} for graph operations, \texttt{sklearn.metrics} for evaluation, \texttt{node2vec 0.4.6} for embeddings, and \texttt{timeit} for computing TTW. To support reproducibility, our code is publicly available at \cite{GraphML-CONTROLLER-AREA-NETWORK-Attack-Detection}.

Section~V-A shows detection results using a baseline model with graph embeddings only, demonstrating significant variation in performance across different \(\omega_t\) and \(\delta_t\) combinations using time-based windows (see Fig.~\ref{fig:auc_roc_embeddings_heatmaps}). Section~V-B evidences how incorporating time series features with graph embeddings enhances detection capabilities, improving AUC-ROC values across various attacks (see Fig.~\ref{fig:auc_roc_embeddings_timeseries_heatmaps}). Section~V-C provides a comparative analysis of detection settings (see Table~\ref{tab:summary_plot}) and also presents TTW as a proxy for computational overhead across attack types (see Table~\ref{tab:ttw_results}). This analysis supports the feasibility of our approach for time-sensitive applications and highlights trade-offs introduced by different window configurations. Section~V-D presents a comparison with state-of-the-art (SOTA) methods, contextualizing our method's performance against existing approaches (see Table~\ref{tab:sota_comparison}). Finally, Section~V-E investigates the effect of using more granular windows based on the number of samples. We show that even under these fine-grained configurations, the framework maintains strong detection performance (see Table~\ref{tab:stat_test_results}).

\subsection{Baseline Model with Graph Embeddings Only}
\begin{figure*}[p]  
    \begin{minipage}{\textwidth}
        \centering
        \includegraphics[width=0.90\textwidth]{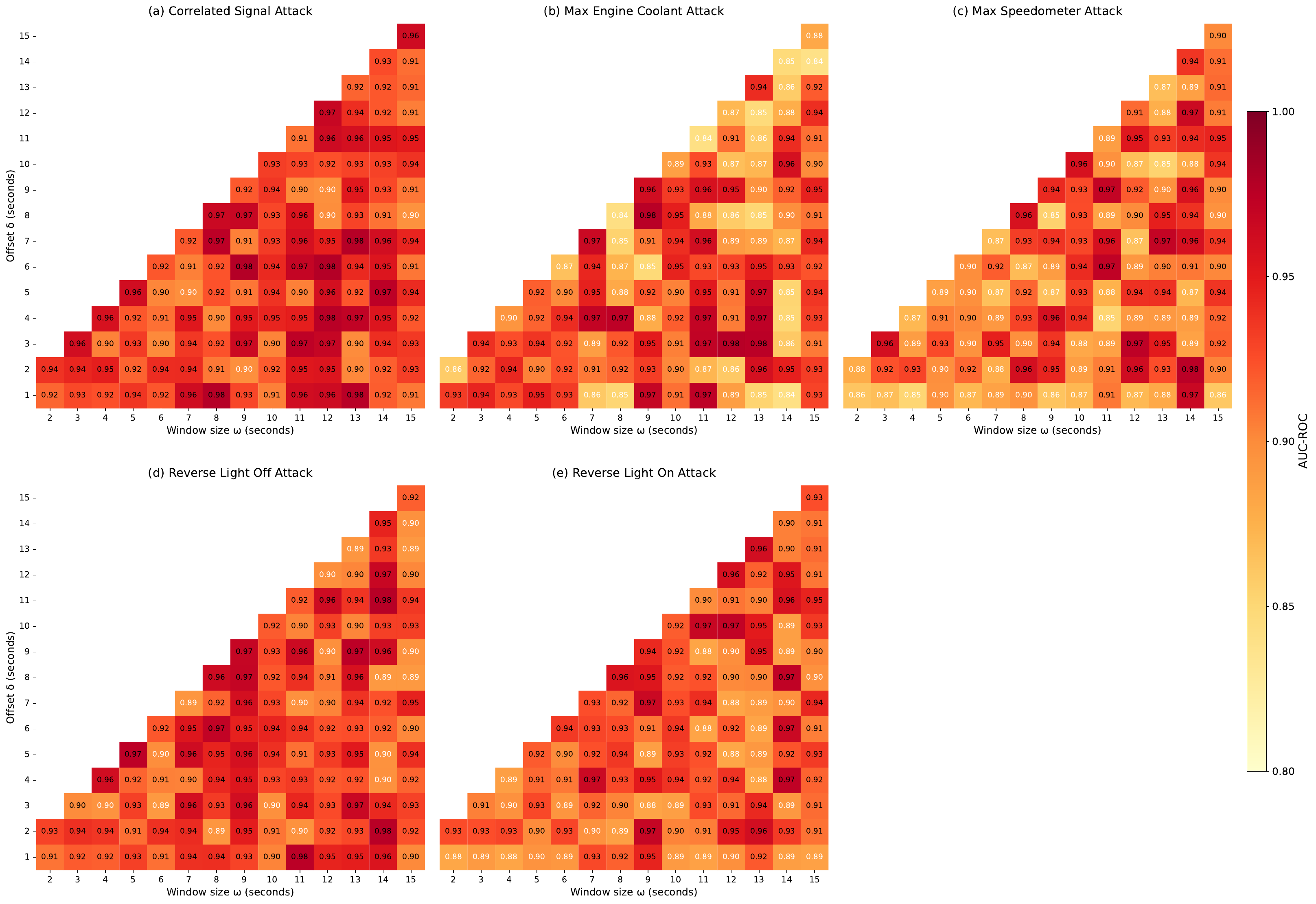}
        \caption{AUC-ROC across different $\omega_t$ and $\delta_t$ combinations for masquerade attacks detected by the baseline model using \emph{only} graph embeddings.}
        \label{fig:auc_roc_embeddings_heatmaps}
    \end{minipage}
    \begin{minipage}{\textwidth}
        \centering
        \includegraphics[width=0.90\textwidth]{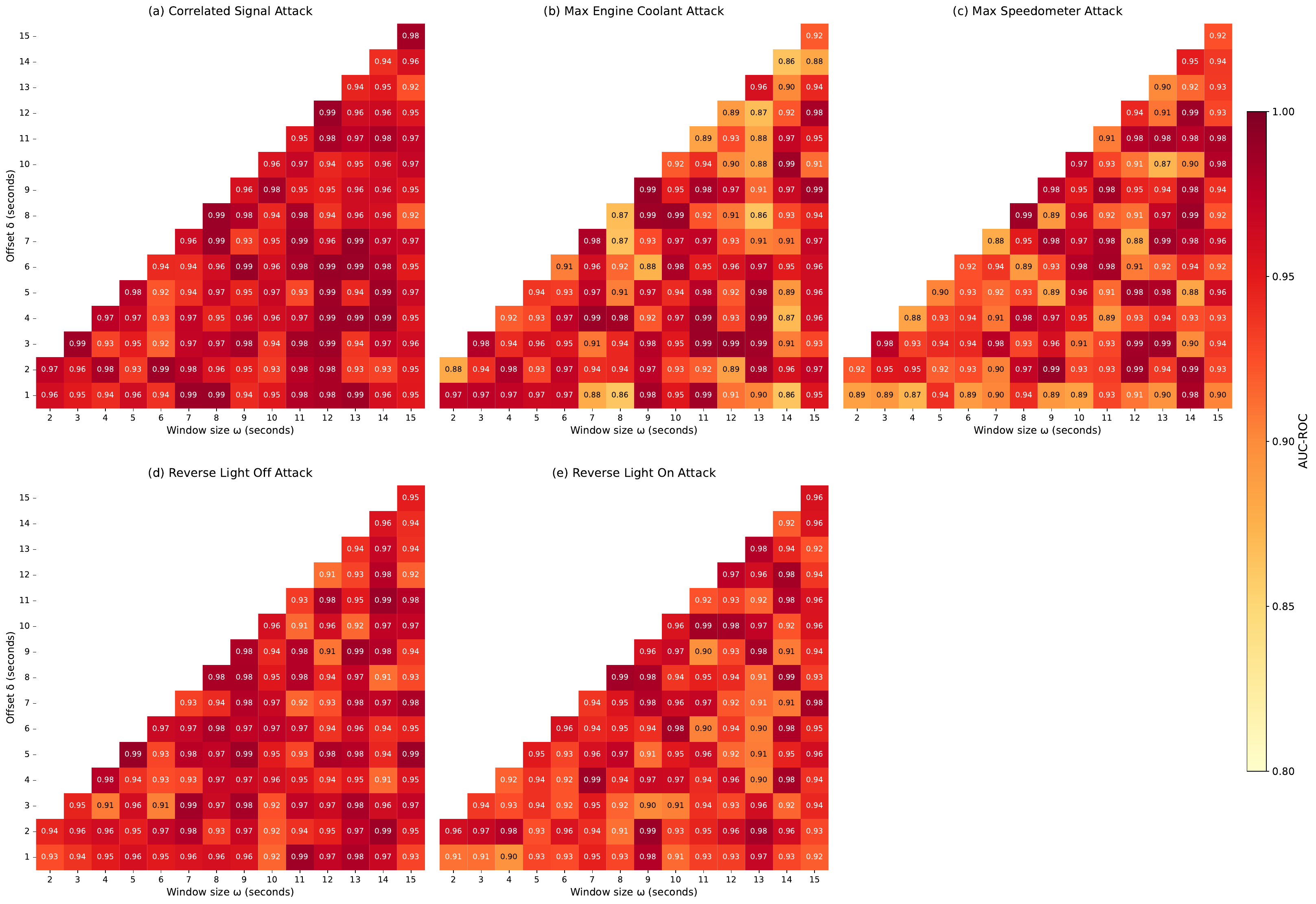}
        \caption{AUC-ROC across different $\omega_t$ and $\delta_t$ combinations for masquerade attacks detected by incorporating \emph{both} graph embeddings and time series features.}
        \label{fig:auc_roc_embeddings_timeseries_heatmaps}
    \end{minipage}
\end{figure*}

Our evaluation begins with the establishment of a baseline model that relies solely on graph embeddings, i.e., focusing only on the topology of the MSGs. This baseline model provides a fundamental comparison point and is essential for assessing the effectiveness of incorporating additional features. Fig.~\ref{fig:auc_roc_embeddings_heatmaps} shows the performance metrics for this baseline serving as the benchmark for our subsequent enhancements. Specifically, the AUC-ROC values for \texttt{correlated\_signal\_attack} reach an optimal performance of 0.98 with a $\omega_t$ of 8 seconds and an $\delta_t$ of 1 second, showcasing strong model effectiveness in this configuration. In contrast, \texttt{reverse\_light\_on\_attack} achieves its highest AUC-ROC of 0.97 with $\omega_t = 7$ seconds and $\delta_t = 4$ seconds, respectively, demonstrating its dependence on extended time frames. In general for the remaining attacks, we notice that high classification results can be obtained for $\omega_t$ and $\delta_t$ of several seconds. These insights confirm that adjusting the configurations to suit different types of attacks is crucial for maximizing detection capabilities. The variance in performance across different settings shows the importance of fine-tuning the sliding window parameters to enhance the efficacy of anomaly detection systems in vehicular networks.

\subsection{Enhanced Model with Graph Embeddings and Time Series}
Building upon our baseline, we incorporate time series features of CAN signals alongside the graph embeddings. This setting harnesses both the structural patterns and the temporal dynamics of CAN frames aiming to achieve a more comprehensive anomaly detection. Fig.~\ref{fig:auc_roc_embeddings_timeseries_heatmaps} depicts the performance metrics of this enhanced model. The inclusion of time series data significantly improves detection capabilities, as evidenced by the AUC-ROC scores. For \texttt{correlated\_signal\_attack}, the model achieves an AUC-ROC peak of 0.99 at a $\omega_t$ of 3 seconds and an $\delta_t$ of 3 seconds. This suggests a strong model response to quick changes in CAN signal patterns. Similarly, \texttt{reverse\_light\_on\_attack} records its highest AUC-ROC at 0.99 for a $\omega_t$ of 7 seconds and an $\delta_t$ of 4 seconds, indicating effective monitoring of critical engine parameters. In general, we notice that when integrating time series features, the required time to get optimal performance is lower than when using only graph embeddings. Specifically, across the five masquerade attacks, the average optimal $\omega_t$ decreases from 9.8 seconds to 6 seconds, while the average optimal $\delta_t$ increases from 3.2 seconds to 4.8 seconds when combining graph embeddings with time series features.

{\color{red}\textbf{Offset Interpretation and Design Rationale:} 
The offset value is not equivalent to a detection delay—rather, it is a mechanism to ensure continuous attack coverage by allowing consecutive windows to overlap. Without this, attacks occurring near the window boundaries could be missed. Increasing the offset (e.g., to 6s) risks skipping attacks entirely, while reducing it (e.g., to 1s) significantly increases computational overhead. Our configuration optimally balances attack coverage and processing feasibility while ensuring the IDS remains responsive.} This 38.8\% reduction in $\omega_t$ (from 9.8\,s to 6\,s) coupled with a 50\% increase in $\delta_t$ (from 3.2\,s to 4.8\,s) allows for more granular analysis of CAN messages while spacing the analyses further apart, enabling our method to capture detailed patterns more effectively.

\color{blue}\begin{table*}[t]
\caption{\color{blue}{COMPARISON OF DETECTION SETTINGS BASED ON AUC-ROC. THE TABLE REPORTS THE MEAN ($\mu$), STANDARD DEVIATION ($\sigma$), MEDIAN ($\eta$), MINIMUM (\textit{min}), AND MAXIMUM (\textit{max}) AUC-ROC VALUES FOR EACH ATTACK SCENARIO. IT ALSO INCLUDES THE AVERAGE (\(\overline{\text{max}}\)) AND STANDARD DEVIATION ($\sigma_{\text{max}}$) OF THE MAXIMUM VALUES OBTAINED ACROSS ALL EXPERIMENT TYPES—GRAPH EMBEDDINGS ONLY, TIME SERIES ONLY, AND THE HYBRID APPROACH.}}
\label{tab:summary_plot}
\color{blue}\centering
\scriptsize

\color{blue}\begin{tabular}{|>{\raggedright\arraybackslash}p{1.5cm}| 
                *{5}{>{\raggedright\arraybackslash}p{2.15cm}|} 
                >{\raggedright\arraybackslash}p{1.6cm}|}

\hline
\textbf{Experiment} &
\textbf{\makecell[t]{\hb{Correlated}\\\hb{signal}}} &
\textbf{\makecell[t]{\hb{Max speedo-}\\\hb{meter}}} &
\textbf{\makecell[t]{\hb{Max engine}\\\hb{coolant}}} &
\textbf{\makecell[t]{\hb{Reverse light}\\\hb{off}}} &
\textbf{\makecell[t]{\hb{Reverse light}\\\hb{on}}} &
\textbf{\makecell[t]{\hb{Experiment}\\\hb{max}}} \\
\hline
\begin{tabular}[t]{@{}l@{}}\hb{Graph} \\ \hb{embeddings} \\ \hb{only}\end{tabular} &
\begin{tabular}[t]{@{}l@{}}
\hb{$\mu = 0.94$} \\
\hb{$\sigma = 0.02$} \\
\hb{$\eta = 0.93$} \\
\hb{min = 0.90 (15,8)} \\
\hb{max = 0.98 (8,1)}
\end{tabular} &
\begin{tabular}[t]{@{}l@{}}
\hb{$\mu = 0.91$} \\
\hb{$\sigma = 0.03$} \\
\hb{$\eta = 0.91$} \\
\hb{min = 0.85 (4,1)} \\
\hb{max = 0.98 (14,2)}
\end{tabular} &
\begin{tabular}[t]{@{}l@{}}
\hb{$\mu = 0.91$}\\
\hb{$\sigma = 0.04$} \\
\hb{$\eta = 0.92$}\\
\hb{min = 0.84 (15,14)} \\
\hb{max = 0.98 (9,8)}
\end{tabular} &
\begin{tabular}[t]{@{}l@{}}
\hb{$\mu = 0.93$}\\
\hb{$\sigma = 0.02$} \\
\hb{$\eta = 0.93$}\\
\hb{min = 0.89 (6,3)} \\
\hb{max = 0.98 (11,1)}
\end{tabular} &
\begin{tabular}[t]{@{}l@{}}
\hb{$\mu = 0.92$}\\
\hb{$\sigma = 0.03$} \\
\hb{$\eta = 0.92$}\\
\hb{min = 0.88 (2,1)} \\
\hb{max = 0.97 (7,4)}
\end{tabular} &
\begin{tabular}[t]{@{}l@{}}
\hb{$\overline{\text{max}} = 0.98$} \\
\hb{$\sigma_{\text{max}} = 0.00$}
\end{tabular} \\
\hline
\begin{tabular}[t]{@{}l@{}}\hb{Time series} \\ \hb{features} \\ \hb{only}\end{tabular} &
\begin{tabular}[t]{@{}l@{}}
\hb{$\mu = 0.82$}\\
\hb{$\sigma = 0.05$} \\
\hb{$\eta = 0.83$}\\
\hb{min = 0.76 (13,7)} \\
\hb{max = 0.90 (10,4)}
\end{tabular} &
\begin{tabular}[t]{@{}l@{}}
\hb{$\mu = 0.85$}\\
\hb{$\sigma = 0.04$} \\
\hb{$\eta = 0.86$}\\
\hb{min = 0.80 (9,6)} \\
\hb{max = 0.91 (11,5)}
\end{tabular} &
\begin{tabular}[t]{@{}l@{}}
\hb{$\mu = 0.79$}\\
\hb{$\sigma = 0.06$} \\
\hb{$\eta = 0.80$}\\
\hb{min = 0.72 (14,9)} \\
\hb{max = 0.89 (12,5)}
\end{tabular} &
\begin{tabular}[t]{@{}l@{}}
\hb{$\mu = 0.81$}\\
\hb{$\sigma = 0.05$} \\
\hb{$\eta = 0.83$}\\
\hb{min = 0.75 (13,8)} \\
\hb{max = 0.88 (9,4)}
\end{tabular} &
\begin{tabular}[t]{@{}l@{}}
\hb{$\mu = 0.84$}\\
\hb{$\sigma = 0.03$} \\
\hb{$\eta = 0.85$}\\
\hb{min = 0.79 (8,5)} \\
\hb{max = 0.90 (10,5)}
\end{tabular} &
\begin{tabular}[t]{@{}l@{}}
\hb{$\overline{\text{max}} = 0.90$} \\
\hb{$\sigma_{\text{max}} = 0.01$}
\end{tabular} \\
\hline
\begin{tabular}[t]{@{}l@{}}\hb{Graph} \\ \hb{embeddings} \\ \hb{+} \\ \hb{time series} \\ \hb{features}\end{tabular} &
\begin{tabular}[t]{@{}l@{}}
\hb{$\mu = 0.96$}\\
\hb{$\sigma = 0.02$} \\
\hb{$\eta = 0.96$}\\
\hb{min = 0.92 (6,3)} \\
\hb{max = 0.99 (3,3)}
\end{tabular} &
\begin{tabular}[t]{@{}l@{}}
\hb{$\mu = 0.94$}\\
\hb{$\sigma = 0.03$} \\
\hb{$\eta = 0.93$} \\
\hb{min = 0.87 (13,10)} \\
\hb{max = 0.99 (8,8)}
\end{tabular} &
\begin{tabular}[t]{@{}l@{}}
\hb{$\mu = 0.94$}\\
\hb{$\sigma = 0.04$} \\
\hb{$\eta = 0.94$}\\
\hb{min = 0.86 (8,1)} \\
\hb{max = 0.99 (7,4)}
\end{tabular} &
\begin{tabular}[t]{@{}l@{}}
\hb{$\mu = 0.96$}\\
\hb{$\sigma = 0.02$} \\
\hb{$\eta = 0.96$}\\
\hb{min = 0.91 (6,3)} \\
\hb{max = 0.99 (5,5)}
\end{tabular} &
\begin{tabular}[t]{@{}l@{}}
\hb{$\mu = 0.94$}\\
\hb{$\sigma = 0.03$} \\
\hb{$\eta = 0.94$}\\
\hb{min = 0.90 (4,1)} \\
\hb{max = 0.99 (7,4)}
\end{tabular} &
\begin{tabular}[t]{@{}l@{}}
\hb{$\overline{\text{max}} = 0.99$} \\
\hb{$\sigma_{\text{max}} = 0.00$}
\end{tabular} \\
\hline
\textbf{\makecell[t]{\hb{Attack} \\ \hb{max}}} &
\begin{tabular}[t]{@{}l@{}}
\hb{$\overline{\text{max}} = 0.98$} \\
\hb{$\sigma_{\text{max}} = 0.01$}
\end{tabular} &
\begin{tabular}[t]{@{}l@{}}
\hb{$\overline{\text{max}} = 0.98$} \\
\hb{$\sigma_{\text{max}} = 0.01$}
\end{tabular} &
\begin{tabular}[t]{@{}l@{}}
\hb{$\overline{\text{max}} = 0.98$} \\
\hb{$\sigma_{\text{max}} = 0.01$}
\end{tabular} &
\begin{tabular}[t]{@{}l@{}}
\hb{$\overline{\text{max}} = 0.98$} \\
\hb{$\sigma_{\text{max}} = 0.01$}
\end{tabular} &
\begin{tabular}[t]{@{}l@{}}
\hb{$\overline{\text{max}} = 0.98$} \\
\hb{$\sigma_{\text{max}} = 0.01$}
\end{tabular} &
\hb{}\\
\hline
\end{tabular}
\end{table*}

\subsection{Metrics Summary}

We compare the effectiveness of the two previously defined configurations and include a time series-only variant as an ablation baseline to assess the individual contribution of temporal features.

Table~\ref{tab:summary_plot} summarizes results across heatmaps. Attack categories are listed in the columns, and detection settings in the rows. Each cell in this table reports summary statistics from the heatmaps for a specific setting and attack type. We focus on metrics including mean (\(\mu\)), standard deviation (\(\sigma\)), median (\(\eta\)), minimum (\(\min\)), and maximum (\(\max\)). Additionally, we compute the average (\(\overline{\max}\)) and standard deviation (\(\sigma_{\max}\)) of the maximum values obtained by each setting and attack type, which are shown in the last row and column of the table.

Graph embeddings combined with time series features consistently outperform both the graph embeddings-only and the time series-only baselines across all attack types. For example, for the \texttt{correlated\_signal\_attack}, the hybrid setting achieves a mean AUC-ROC (\(\mu\)) of 0.96 with a standard deviation (\(\sigma\)) of 0.02, compared to 0.94 and 0.02 for the graph-only baseline, and 0.82 and 0.05 for the time-series-only baseline. Similarly, for \texttt{max\_engine\_coolant\_attack}, the hybrid setting achieves a mean AUC-ROC of 0.94 and standard deviation of 0.04, versus 0.91 and 0.04 for graph-only, and 0.79 and 0.06 for time-series-only. A comparable improvement is seen across \texttt{max\_speedometer\_attack} (0.94 vs. 0.91 vs. 0.85), \texttt{reverse\_light\_off\_attack} (0.96 vs. 0.93 vs. 0.81), and \texttt{reverse\_light\_on\_attack} (0.94 vs. 0.92 vs. 0.84), with all standard deviations remaining within similar bounds.

\textit{How do spoofed frames differ in feature domains?}

In feature space, spoofed frames differ from benign traffic primarily through anomalies in structural transitions and temporal regularities \cite{moriano2024benchmarkingunsupervisedonlineids}. Our evaluation shows that the fused feature representation enables near-perfect separation between benign and malicious traffic. Specifically, our hybrid model achieves a mean AUC-ROC of 0.99 across all masquerade attacks, indicating that the combined structural and temporal deviations are sufficient for discrimination, even when content and timing are superficially plausible. Notably, difficult-to-detect attacks such as \texttt{max\_engine\_coolant}, previously noted for weak signal-level deviation, are still effectively detected by our approach. This supports the conclusion that our hybrid features amplify subtle inconsistencies—particularly in message transition structure and short-term temporal signal patterns—that standalone models may miss. Prior CAN intrusion studies also suggest that such composite feature sets yield superior discrimination~\cite{cho2016fingerprinting, tariq2020cantransfer}.

We now summarize median and maximum statistics for each configuration.
In \texttt{correlated\_signal\_attack}, the hybrid setting yields a median (\(\eta\)) of 0.96, a minimum (\(\min\)) of 0.92, and a maximum (\(\max\)) of 0.99. These statistics capture the central tendency and spread of detection performance across all configurations. Notably, the time-series-only baseline consistently underperforms in terms of both average and peak metrics, which further validates the necessity of combining temporal and structural features for detecting stealthy attacks. Moreover, we observe that for \texttt{max\_engine\_coolant\_attack}, the average maximum AUC-ROC (\(\overline{\max}\)) reaches 0.99 for the hybrid model, 0.98 for the graph-only baseline, and only 0.89 for the time-series-only variant, with a standard deviation of (\(\sigma_{\max}\)) = 0.00 or 0.01—highlighting both the performance gap and stability offered by the hybrid method across experimental trials.

{\color{red}
To provide clearer insights into the IDS’s classification performance, we computed the false positive rate (FPR) and false negative rate (FNR) at a selected operating threshold derived from the receiver operating characteristic (ROC) curve~\cite{FAWCETT2006861}. Specifically, we selected the threshold that maximizes Youden’s \(J\) index ($J = \text{TPR} - \text{FPR}$), a standard method for identifying the optimal balance between true positive and false positive rates in binary classification tasks~\cite{yang2025unflows}. This ensures a meaningful trade-off between sensitivity and specificity, supporting practical deployment. Table~\ref{tab:fpr_fnr} reports the average FPR and FNR across all attack types and $(\omega_t, \delta_t)$ combinations, highlighting consistent improvements—especially in FNR—when time series features are integrated.
}

{\color{red}\begin{table}[h]
\color{red}\centering
\caption{\color{red}MEAN AND STANDARD DEVIATION OF FPR AND FNR AT THE OPTIMAL ROC-DERIVED THRESHOLD (YOUDEN’S \(J\) INDEX). ``GRAPH EMBEDDINGS'' REFERS TO THE BASELINE; ``+ TIME SERIES FEATURES'' DENOTES THE ENHANCED SETTING.}

\label{tab:fpr_fnr}
\begin{adjustbox}{max width=0.5\textwidth}
\begin{tabular}{@{}llcc@{}}
\toprule
\textbf{Attack Name} & \textbf{Version} & \textbf{FPR} & \textbf{FNR} \\
 & & \textbf{(mean $\pm$ std)} & \textbf{(mean $\pm$ std)} \\
\midrule

\texttt{correlated\_signal} 
 & Graph embeddings & 0.04 $\pm$ 0.01 & 0.09 $\pm$ 0.02 \\ 
 & + time series features & 0.02 $\pm$ 0.01 & 0.07 $\pm$ 0.02 \\
\hdashline

\texttt{max\_speedometer} 
 & Graph embeddings & 0.01 $\pm$ 0.01 & 1.20 $\pm$ 0.22 \\ 
 & + time series features & 0.00 $\pm$ 0.01 & 0.80 $\pm$ 0.17 \\
\hdashline

\texttt{max\_engine\_coolant} 
 & Graph embeddings & 0.02 $\pm$ 0.01 & 0.27 $\pm$ 0.05 \\ 
 & + time series features & 0.01 $\pm$ 0.01 & 0.21 $\pm$ 0.05 \\
\hdashline

\texttt{reverse\_light\_off} 
 & Graph embeddings & 0.06 $\pm$ 0.01 & 0.01 $\pm$ 0.01 \\ 
 & + time series features & 0.04 $\pm$ 0.01 & 0.01 $\pm$ 0.01 \\
\hdashline

\texttt{reverse\_light\_on} 
 & Graph embeddings & 0.01 $\pm$ 0.01 & 0.87 $\pm$ 0.19 \\ 
 & + time series features & 0.00 $\pm$ 0.01 & 0.69 $\pm$ 0.14 \\

\bottomrule
\end{tabular}
\end{adjustbox}
\end{table}
}

{\color{red}
Table~\ref{tab:ttw_results} summarizes TTW results, including mean, median, standard deviation, minimum, and maximum processing times, with corresponding optimal $\delta_t$ and $\omega_t$ combinations shown in parentheses. Table~\ref{tab:ttw_results} demonstrates that our hybrid approach achieves reasonable computational overhead across attack types. For instance, the \texttt{correlated\_signal\_attack} achieves a mean TTW of 2.36s, with a best-case performance as low as 0.024s, highlighting feasibility for real-time application with further optimization.
}

{\color{red}\begin{table}[h]
\color{red}\centering
\caption{\color{red}TTW SUMMARY ACROSS ATTACK TYPES WITH MIN/MAX CONFIGURATIONS.}

\label{tab:ttw_results}
\begin{adjustbox}{max width=0.5\textwidth}
\begin{tabular}{@{}lcccll@{}}
\toprule
\textbf{Attack Name} & \textbf{Mean (s)} & \textbf{Std (s)} & \textbf{Median (s)} & \textbf{Min TTW (s)} & \textbf{Max TTW (s)} \\
\midrule

\texttt{correlated\_signal} 
& 2.36 & 1.21 & 2.44 
& 0.024 (12.0, 1.0) 
& 6.84 (10.0, 1.0) \\
\hdashline

\texttt{max\_speedometer} 
& 2.67 & 1.29 & 2.65 
& 0.005 (13.0, 4.0) 
& 20.82 (13.0, 4.0) \\
\hdashline

\texttt{max\_engine\_coolant} 
& 2.30 & 1.16 & 2.30 
& 0.237 (10.0, 5.0) 
& 4.89 (15.0, 9.0) \\
\hdashline

\texttt{reverse\_light\_off} 
& 2.60 & 1.29 & 2.59 
& 0.027 (10.0, 1.0) 
& 6.22 (15.0, 14.0) \\
\hdashline

\texttt{reverse\_light\_on} 
& 2.80 & 1.36 & 2.93 
& 0.005 (13.0, 1.0) 
& 8.53 (9.0, 2.0) \\

\bottomrule
\end{tabular}
\end{adjustbox}
\end{table}
}

{\color{blue} Table~\ref{tab:auc_ttw_tradeoff} presents the AUC-ROC and corresponding TTW for selected $(\omega_t, \delta_t)$ configurations across all attack types. Each row reflects a configuration that balances high detection accuracy with practical runtime cost. Note that for \texttt{reverse\_light\_on}, the second row highlights a lower-latency alternative with a slight reduction in AUC-ROC, illustrating the effect of window size and offset on real-time feasibility.
We show that our hybrid approach consistently achieves near-perfect detection (0.99 AUC-ROC) across all masquerade attacks, while also revealing the varying computational overhead associated with different optimal configurations. For instance, the \texttt{correlated\_signal} attack is detected with 0.99 AUC-ROC at 0.44s TTW, highlighting its efficiency. This comprehensive view of performance and runtime is critical for deploying IDS in resource-constrained vehicular environments.}

\color{blue}\begin{table}[h]
\color{blue}\centering
\caption{\color{blue}TRADE-OFF BETWEEN DETECTION ACCURACY AND RUNTIME EFFICIENCY FOR SELECTED $(\omega_t, \delta_t)$ CONFIGURATIONS.}

\label{tab:auc_ttw_tradeoff}
\begin{adjustbox}{max width=\textwidth}
\color{blue}\begin{tabular}{@{}lcccc@{}}
\toprule
\textbf{Attack Name} & \textbf{$\omega_t$ (s)} & \textbf{$\delta_t$ (s)} & \textbf{AUC-ROC} & \textbf{TTW (s)} \\
\midrule
\texttt{correlated\_signal}      & 3 & 3 & 0.99 & 0.44 \\
\texttt{max\_speedometer}        & 8 & 8 & 0.99 & 1.85 \\
\texttt{max\_engine\_coolant}    & 7 & 4 & 0.99 & 0.92 \\
\texttt{reverse\_light\_off}     & 5 & 5 & 0.99 & 0.73 \\
\multirow{2}{*}{\texttt{reverse\_light\_on}}    
                                 & 7 & 4 & 0.99 & 1.10 \\
                                 & 4 & 2 & 0.96 & 0.18 \\
\bottomrule
\end{tabular}
\end{adjustbox}
\end{table}

\begin{figure}[h]
    \centering
    \includegraphics[width=0.85\linewidth]{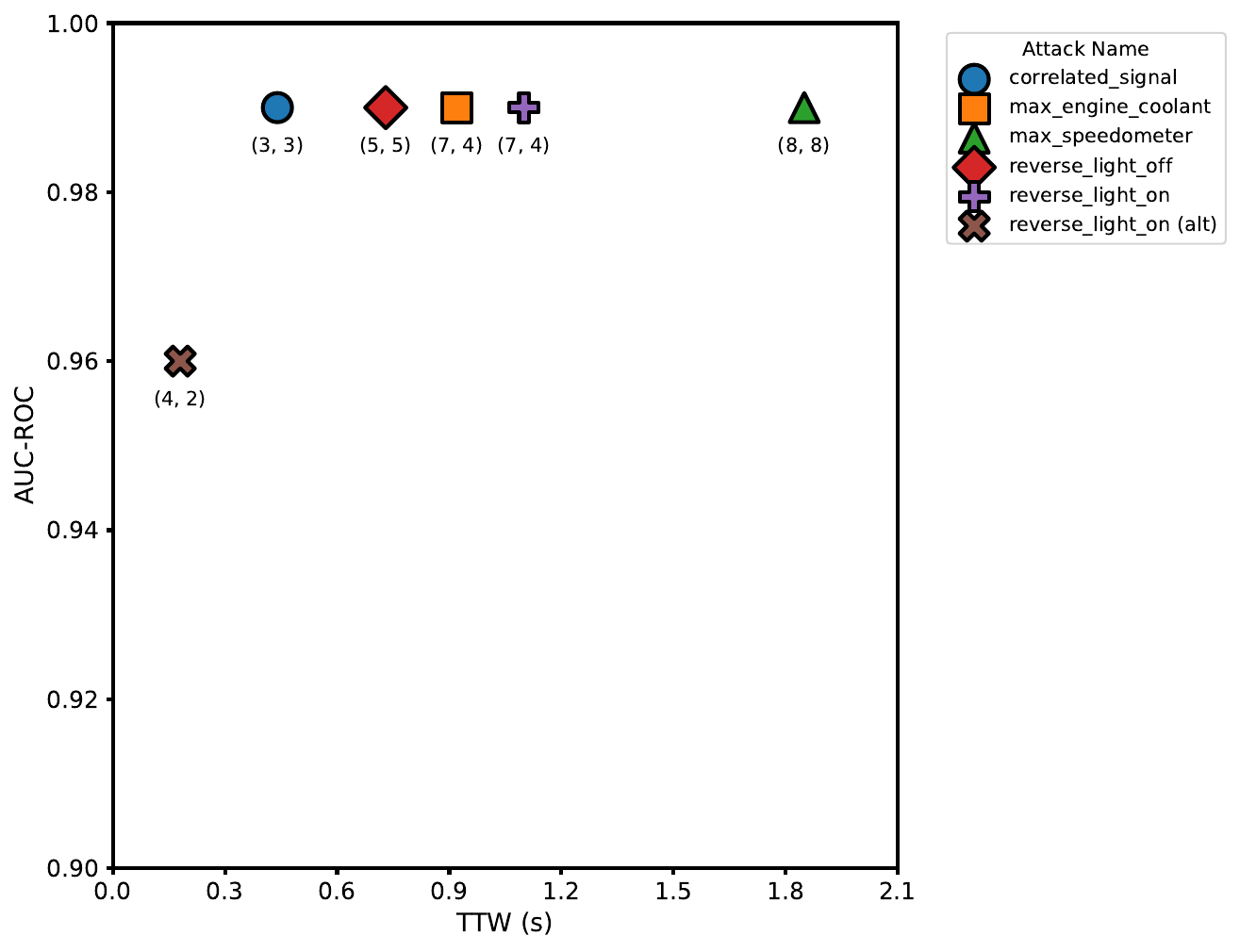}
    \caption{\color{blue}Trade-off between detection accuracy (AUC-ROC) and runtime cost (TTW) for selected $(\omega_t, \delta_t)$ configurations. Values in parentheses below each point indicate the $\omega_t$ and $\delta_t$ used in that configuration. The \texttt{reverse\_light\_on (alt)} point reflects an alternate low-latency configuration for the same attack, included to illustrate the practical trade-off.}
    \label{fig:ttw_tradeoff}
\end{figure}

{\color{blue} Figure~\ref{fig:ttw_tradeoff} visualizes this trade-off by plotting AUC-ROC against TTW, annotated with the corresponding $(\omega_t, \delta_t)$ values. While most configurations achieve near-identical AUC-ROC scores (0.99), the associated TTW varies, reflecting the computational impact of wider temporal contexts. Note that the alternate \texttt{reverse\_light\_on} configuration, for example, reduces TTW from 1.10\,s to 0.18\,s with only a minor drop in AUC-ROC. We show the importance of selecting $(\omega_t, \delta_t)$ pairs that offer an optimal balance between detection accuracy and inference latency.}

\color{black} To further validate the differences in detection effectiveness of both settings, we employed the Mann-Whitney \(\mathrm{U}\) test and the Kolmogorov-Smirnov \((\mathrm{KS}\)) test. The Mann-Whitney \(\mathrm{U}\) test \cite{mann1947test} is a non-parametric statistical test that evaluates whether there is a difference between two independent samples. The \(\mathrm{KS}\) test
 \cite{berger2014kolmogorov} is another non-parametric test that assesses whether two samples come from the same distribution. In our analysis, we compared the AUC-ROC value distributions of the two settings. Here, the null hypothesis is that the AUC-ROC values for the setting combining embeddings with time series features are less than or equal to those for the setting using only embeddings, while the alternative hypothesis is that the AUC-ROC values for the setting combining embeddings with time series features are greater than those for the setting using only embeddings. A low \( p \)-value indicates a significant difference between these two settings. We focused on a significance level of 0.05. 

 Table \ref{tab:test_results} shows significant differences between the observed distributions for all masquerade attacks. In each of the attacks, the Mann-Whitney U and \(\mathrm{KS}\) tests produce low \( p \)-values, indicating a significant deviation from the expected distribution. This supports our hypothesis of relying on annotated MSGs to effectively detect masquerade attacks at various windows and offset combinations. Consequently, we reject the null hypothesis for all tested masquerade attacks, confirming that the combined setting performs significantly better.

\begin{table}[h]
\centering
\caption{MANN-WHITNEY U AND KOLMOGOROV-SMIRNOV TEST RESULTS FOR TIME-BASED WINDOWS.}

\label{tab:test_results}
\begin{adjustbox}{max width=0.50\textwidth}
\begin{tabular}{@{}lcccl@{}}
\toprule
\textbf{Attack Name} & \textbf{U Statistic} & \textbf{$p$-value} & \textbf{KS Statistic} & \textbf{$p$-value} \\
\midrule
\texttt{correlated\_signal} & 11337.0 & 5.55e-16 & 0.46 & 3.81e-12 \\
\hdashline
\texttt{max\_speedometer} & 10082.0 & 7.98e-09 & 0.33 & 2.34e-06 \\
\hdashline
\texttt{max\_engine\_coolant} & 9864.0 & 8.01e-08 & 0.33 & 2.34e-06 \\
\hdashline
\texttt{reverse\_light\_off} & 11033.0 & 4.98e-14 & 0.43 & 1.73e-10 \\
\hdashline
\texttt{reverse\_light\_on} & 10818.0 & 9.84e-13 & 0.43 & 1.73e-10 \\
\bottomrule
\end{tabular}
\end{adjustbox}
\end{table}

{\color{blue}\subsection{Comparison with SOTA Methods:}
We compare our hybrid approach against two relevant SOTA baselines on the ROAD dataset to further contextualize our method's performance. Our proposed hybrid method integrates structural graph features with temporal dynamics to enhance anomaly detection. SOTA baselines include:

\textbf{Moriano et al.~\cite{moriano2024benchmarkingunsupervisedonlineids}} --- An unsupervised time series approach which is evaluated on the ROAD dataset \cite{verma2022addressing} and reports AUC-ROC and TTW using sliding window parameters $(\omega_t, \delta_t)$ aligning with our evaluation design. We replicated their best-performing configuration and included results for each masquerade attack.

\textbf{CANShield~\cite{shahriar2022canshield}} --- A deep learning-based method that operates on signal-level representations, applied to the ROAD dataset~\cite{verma2022addressing}. While effective in detecting various attack classes, it does not report TTW nor explicitly define $(\omega_t, \delta_t)$ values. Instead, it segments data internally using \texttt{window\_step\_test} parameters. Additionally, due to the high-dimensional signal-level input and deeper architecture, the approach is considerably more computationally intensive on ROAD. As such, TTW is omitted and only AUC-ROC is reported for this baseline.

Table~\ref{tab:sota_comparison} summarizes the attack-wise AUC-ROC and TTW scores. Our approach achieves the highest AUC-ROC (0.99) across all five masquerade attacks while consistently demonstrating lower TTW per window. The optimal $(\omega_t, \delta_t)$ values for our configuration are presented in Table~\ref{tab:auc_ttw_tradeoff}. For Moriano et al. \cite{moriano2024benchmarkingunsupervisedonlineids}, the values correspond to their best-performing public configuration. We show the effectiveness and practicality of our method for real-time detection, highlighting its robust performance under constrained latency budgets. The code to reproduce these baselines has been added to our GitHub repository \cite{GraphML-CONTROLLER-AREA-NETWORK-Attack-Detection}.

\begin{table}[h]
\color{blue}\centering
\caption{\color{blue}COMPARATIVE PERFORMANCE OF OUR HYBRID APPROACH VERSUS SOTA METHODS ON THE ROAD DATASET. AUC-ROC AND TTW VALUES CORRESPOND TO EACH METHOD’S BEST-PERFORMING CONFIGURATION PER ATTACK.}
\label{tab:sota_comparison}
\begin{adjustbox}{max width=0.5\textwidth} 
\begin{tabular}{@{}lcccc@{}}
\toprule
\textbf{Attack Name} & \textbf{Metric} & \textbf{Ours} & \textbf{\makecell{Moriano}} & \textbf{\makecell{CANShield\\}} \\
\midrule
\multirow{2}{*}{\texttt{correlated\_signal}}
& AUC-ROC & \textbf{0.99} & 0.96 & 0.93 \\
& TTW (s) & 0.44 & 0.67 & N/A \\
\hdashline 
\multirow{2}{*}{\texttt{max\_speedometer}}
& AUC-ROC & \textbf{0.99} & 0.97 & 0.94 \\
& TTW (s) & 1.85 & 2.01 & N/A \\
\hdashline
\multirow{2}{*}{\texttt{max\_engine\_coolant}}
& AUC-ROC & \textbf{0.99} & 0.94 & 0.91 \\
& TTW (s) & 0.92 & 1.04 & N/A \\
\hdashline
\multirow{2}{*}{\texttt{reverse\_light\_off}}
& AUC-ROC & \textbf{0.99} & 0.95 & 0.92 \\
& TTW (s) & 0.73 & 0.89 & N/A \\
\hdashline
\multirow{2}{*}{\texttt{reverse\_light\_on}}
& AUC-ROC & \textbf{0.99} & 0.93 & 0.90 \\
& TTW (s) & 1.10 & 1.36 & N/A \\
\midrule 
\textbf{Average}
& AUC-ROC & \textbf{0.99} & 0.95 & 0.92 \\
& TTW (s) & \textbf{0.97} & 1.19 & N/A \\
\bottomrule
\end{tabular}
\end{adjustbox}
\vspace{0.5em}
\footnotesize\noindent\parbox{0.5\textwidth}{$^\dagger$\textit{CANShield does not explicitly define $(\omega_t, \delta_t)$ values but adopts an equivalent segmentation strategy using \texttt{window\_step\_test} parameters; TTW is omitted as its framework does not report latency per window.}}
\end{table}}

\subsection{Evaluation Based on Sample Windows}
We investigate the effect of sample-based windows in our detection framework's performance, as opposed to time-based windows. Our aim is to understand the impact of varying the number of CAN message samples in each sliding window on masquerade attack detection. In this context, time-based windows refer to windows defined by a fixed duration, whereas sample-based windows are defined by a fixed number of CAN messages. We experiment with different $\omega_s$ and $\delta_s$ to find the optimal configurations for detecting anomalies.

Due to space constraints, we do not include detailed heatmap results in this version of the paper. However, note that our sample-based results suggest that the integration of time series features with graph embeddings enhances the model's ability to detect anomalies, even when requiring slightly larger sample windows. {\color{red}The improvement in detection capability comes with a measurable increase in computational requirements, as quantified by the TTW metrics in Section~V-C. Further optimization and hardware-specific tuning will be necessary to ensure real-time feasibility in production vehicles.}

We also found the distributions of AUC-ROC values to be significantly different between the two settings. The Mann-Whitney \(\mathrm{U}\) and \(\mathrm{KS}\) tests confirmed this difference, indicating the superior detection performance of the combined setting. Table~\ref{tab:stat_test_results} summarizes the results, showing significant differences across all tested masquerade attacks. For instance, \texttt{correlated\_signal\_attack} yielded a Mann-Whitney U statistic of 1069.5 with a \( p \)-value of 9.84e-07, and a KS statistic of 0.58 with a \( p \)-value of 2.61e-06. Similar trends were observed for other attacks, such as \texttt{max\_engine\_coolant\_attack} and \texttt{max\_speedometer\_attack}, confirming the effectiveness of incorporating time series features with graph embeddings. The results for each attack type consistently indicate that the combined setting outperforms the embeddings-only setting.

\begin{table}[h]
\centering
\caption{MANN-WHITNEY U AND KOLMOGOROV-SMIRNOV TEST RESULTS FOR SAMPLE-BASED WINDOWS.}

\label{tab:stat_test_results}
\begin{adjustbox}{max width=0.50\textwidth}
\begin{tabular}{@{}lcccl@{}}
\toprule
\textbf{Attack Name} & \textbf{U Statistic} & \textbf{$p$-value} & \textbf{KS Statistic} & \textbf{$p$-value} \\
\midrule
\texttt{correlated\_signal} & 1069.5 & 9.84e-07 & 0.58 & 2.61e-06 \\
\hdashline
\texttt{max\_speedometer} & 1085.0 & 4.20e-07 & 0.56 & 9.30e-06 \\
\hdashline
\texttt{max\_engine\_coolant} & 811.5 & 0.033 & 0.33 & 0.018 \\
\hdashline
\texttt{reverse\_light\_off} & 1040.5 & 4.84e-06 & 0.50 & 9.36e-05 \\
\hdashline
\texttt{reverse\_light\_on} & 1112.0 & 8.35e-08 & 0.61 & 6.75e-07 \\
\bottomrule
\end{tabular}
\end{adjustbox}
\end{table}


\section{Discussion} \label{sec:Discussion}

This paper presents a novel approach for detecting masquerade attacks in CAN using graph ML. By combining graph embeddings with time series features, our method significantly improves detection performance across all masquerade attack types in the ROAD dataset. These gains are statistically validated using the Mann-Whitney \(\mathrm{U}\) and \(\mathrm{KS}\) tests (see Tables~\ref{tab:test_results} and~\ref{tab:stat_test_results}). For instance, detection of the \texttt{max\_engine\_coolant\_attack}—previously difficult to detect due to weak signal deviation \cite{moriano2024benchmarkingunsupervisedonlineids}—was notably improved in the enhanced setting, demonstrating stronger anomaly sensitivity. Our evaluation also shows that using sample-based windows—particularly with smaller \(\omega\) values—enables the detection of subtle anomalies via fine-grained analysis.

{\color{blue}\textit{\textbf{What specific deviations does the model capture?}}\\
Our hybrid framework captures subtle, cumulative deviations that even sophisticated spoofing cannot perfectly hide. We leverage two complementary domains:
\begin{itemize}[label={}]
    \item \textbf{Structural deviations (via MSGs and graph embeddings):} Masquerade attacks involve an attacker suppressing a legitimate ECU and impersonating it. Attackers cannot perfectly replicate intricate real-time inter-message dependencies, message orderings, and complex sequential patterns (graph motifs) characteristic of benign ECU communication. Prior work demonstrates motif frequency distributions as sensitive indicators of anomalies~\cite{milo2002network}. Graph-based approaches effectively capture these structural anomalies in CAN traffic. Our approach leverages \textit{node2vec} to embed these deviations, extracting low-dimensional representations that preserve the complex topology of MSGs.
    \item \textbf{Temporal and content deviations (via timeseries features):} Masquerade attacks can introduce subtle inconsistencies in physical signal values or precise timing patterns, even if overall frequency is matched. Decoded signal values may show anomalies in statistical properties (mean, standard deviation) or inter-signal correlations, which are stable in normal driving but deviate under spoofing~\cite{zhang2020cansec}. Our node annotation technique enriches CAN ID nodes with these statistics, enabling detection of nuanced temporal and content-level inconsistencies. Similar approaches prove effective for high-frequency signal perturbations in automotive IDS~\cite{Zhang2023, moriano2024benchmarkingunsupervisedonlineids}.
\end{itemize}
The strength of our hybrid approach lies in its ability to fuse these structural graph embeddings with the temporal signal statistics. A deviation missed by one modality might be detected by the other, providing a robust, multi-faceted detection capability against attacks that blend seamlessly into normal traffic.}

The effectiveness of our approach highlights the value of combining structural and temporal insights in CAN IDS. Graph embeddings capture ECU interactions, while time series features reflect evolving signal patterns. This dual analysis improves detection and allows more efficient sliding window configuration—reducing the optimal offset \(\delta_t\) by 38.8\% (9.8\,s to 6\,s) and increasing the window length \(\omega_t\) by 50\% (3.2\,s to 4.8\,s). These adjustments enhance timely detection and emphasize the importance of securing CAN-based systems as vehicle architectures evolve.

{\color{red}
While automotive Ethernet adoption grows (projected to reach 11.6 billion US dollars by 2032 \cite{GMI2024}, it primarily serves high-bandwidth domains like ADAS and infotainment (38\% market share in 2023). However, CAN remains critical for safety subsystems (e.g., engine, braking) due to its role in hybrid architectures and persistent vulnerabilities. Over 90\% of modern vehicles use hybrid CAN/Ethernet networks, with CAN handling safety-critical low-speed data (powertrain/chassis) while Ethernet supplements infotainment \cite{GMI2024}. CAN’s simplicity and cost ensure its ubiquity in existing fleets, with Ethernet adoption concentrated in high-end models. Despite Ethernet’s growth, CAN lacks encryption/authentication, making it vulnerable to attacks like brake disablement via OBD-II injection \cite{SealingTech2024}. These risks persist even in hybrid architectures, as compromised Ethernet gateways can inject malicious CAN messages. Our work addresses these gaps with lightweight CAN-specific detection mechanisms. CAN security remains foundational, as even Ethernet-reliant systems interface with CAN for critical functions.
}

{\color{red}
Our work distinguishes itself from prior methods by integrating graph-based and time series analysis to address the subtle nature of masquerade attacks. While existing approaches focus on single-modality detection, our hybrid framework captures both structural and temporal anomalies, enabling the detection of sophisticated intrusions that traditional IDS might miss. This dual approach not only improves detection accuracy but also provides a scalable solution adaptable to the dynamic nature of modern vehicular networks.
}

{\color{red}
Our parameter choices are empirically grounded in the ROAD dataset’s attack profiles, balancing accuracy and practicality. While shorter windows may suit other threat models, masquerade detection necessitates analyzing sustained patterns. For safety-critical functions requiring sub-second detection (e.g., brake disengagement), adaptive parameter tuning and hardware-specific optimizations will be explored in future work.
}{\color{red}While our approach demonstrates promising detection capabilities, its real-time deployment in resource-constrained vehicular environments requires further validation on embedded platforms (e.g., ARM-based boards). Our current results, based on realistic CAN logs generated on a physical vehicle, focus on validating the detection methodology rather than finalizing an on-vehicle implementation.}

{\color{red}While cryptographic measures such as encryption and MACs are effective defenses against masquerade attacks, our anomaly detection framework provides an additional layer of security. By detecting subtle deviations in CAN traffic patterns, our approach complements cryptographic defenses, ensuring robust protection against sophisticated attacks, including those that bypass cryptographic approaches.}

While our results are promising, it is important to acknowledge the limitations of this study:

\textbf{Lack of Validation in a Real Environment:} Our study is simulation-based and lacks validation in real vehicle environments, where factors like latency, sensor noise, and varying driving conditions could impact performance. Future work should involve empirical validation on a moving vehicle, including hardware integration and performance tuning.

\textbf{Evaluation Based Solely on ROAD Dataset:} While comprehensive, the ROAD dataset \cite{verma2022addressing} may not capture the full range of real-world attack scenarios. Future studies should incorporate additional datasets with diverse and recent attack patterns to improve robustness and generalizability.

\textbf{Computational Complexity:} The graph embedding phase may introduce computational overhead. Future work should explore algorithmic optimization and hardware acceleration for improved efficiency.

\textbf{Use of Default Parameter Values in Detection Algorithms:} This study relied on default parameters in detection algorithms, which may not yield optimal performance. Future research should involve extensive hyperparameter tuning to maximize detection accuracy in application-specific contexts.

{\color{blue}\textbf{Limits of Detectability:}
No system is impervious to highly sophisticated attackers. The detectability limits for our framework are as follows:
\begin{itemize}[label={}]
    \item \textbf{Perfect mimicry:} If an attacker can replicate the full message transition topology and multivariate signal relationships of a legitimate ECU—including physical-layer jitter and temporal correlations—the detection margin narrows. However, this level of mimicry requires access to proprietary ECU timing, bus arbitration, and signal generation logic, which is extremely difficult without insider knowledge~\cite{Miller2014RemoteAttack}.
    \item \textbf{Short or bursty attacks:} Transient attacks that operate below the minimum window size or inject very few anomalous transitions may evade detection. This is a known limitation in time-window-based IDS designs~\cite{Jedh2021}.
    \item \textbf{Overlapping statistical distributions:} In rare cases, spoofed messages may fall within the natural variability of benign traffic, making them indistinguishable even in a fused feature space. Defense against this class of attack may require incorporating physical-layer fingerprints~\cite{cho2016fingerprinting} or entropy-based signal modeling~\cite{Marchetti2017}.
\end{itemize}
    
\color{black}In summary, our study contributes to the growing body of research on graph-based CAN IDS by demonstrating the effectiveness of combining graph embeddings with time series features. This approach not only enhances detection capabilities but also provides a flexible framework adaptable to various vehicular systems and attack patterns.

\section{Conclusion} \label{sec:Conclusion}

This paper presents a novel framework for detecting masquerade attacks in CANs by integrating graph ML with time series analysis. By representing CAN frames as MSGs and enriching each node with contextual statistical features, the proposed approach captures both the structural and temporal dynamics of in-vehicle communication. Experiments on the ROAD dataset \cite{verma2022addressing} show that this hybrid model consistently outperforms graph-only baselines across all tested masquerade attack types. These improvements are statistically validated through the Mann-Whitney \(\mathrm{U}\) and Kolmogorov-Smirnov tests (see Tables~\ref{tab:test_results} and~\ref{tab:stat_test_results}), confirming the robustness of our detection capabilities across various attack scenarios.

Future work will explore deploying this framework in online detection settings, investigate lightweight graph-based architectures such as those introduced by Zhang et al.~\cite{Zhang2023}, and assess its adaptability across diverse datasets. We also aim to explore integration with foundation models for enhancing generalization in evolving vehicular environments.

\section*{Acknowledgments} \label{sec:Acknowledgments}

This manuscript has been authored by UT-Battelle, LLC under Contract No. DE-AC05-00OR22725 with the U.S. Department of Energy. The publisher, by accepting the article for publication, acknowledges that the U.S. Government retains a non-exclusive, paid up, irrevocable, world-wide license to publish or reproduce the published form of the manuscript, or allow others to do so, for U.S. Government purposes. The DOE will provide public access to these results in accordance with the DOE Public Access Plan (\url{http://energy.gov/downloads/doe-public-access-plan}). This research was sponsored in part by Oak Ridge National Laboratory’s (ORNL’s) Laboratory Directed Research and Development program through the Sustainable Research Pathways (SRP) program and by the DOE. This research is also partially funded by US Department of Energy Award No. DE-FE0032089. There was no additional external funding received for this study. The funders had no role in study design, data collection and analysis, decision to publish, or preparation of this manuscript.


\small
\bibliographystyle{IEEEtran}
\bibliography{90-bibliography}

\IEEEtriggeratref{56}
\balance

\begin{IEEEbiography}[{\includegraphics[width=1in,height=1.25in,clip,keepaspectratio]{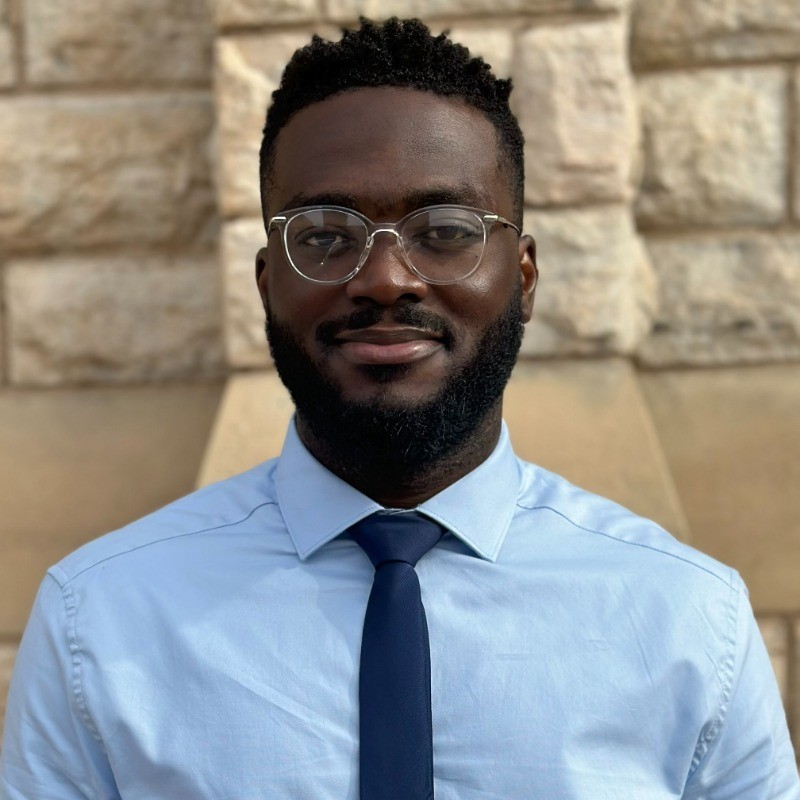}}]{William Marfo}
(Member, IEEE) received the Ph.D. degree in Computer Science from the University of Texas at El Paso, USA. He is currently an Associate Specialist and Cyberinfrastructure Software Engineer with the Cascadia Region Earthquake Science Center (CRESCENT) at the University of California, Davis, USA. His research interests include machine learning, high-performance computing, and cybersecurity for complex computational systems. He develops scalable software and AI/ML models for large-scale seismic data processing and interdisciplinary earthquake science. He previously conducted research at U.S. Department of Energy national laboratories and was named a 2025 Trusted CI Fellow by the US National Science Foundation (NSF) Cybersecurity Center of Excellence. Dr. Marfo is a member of ACM, SIAM, AMS, and ISACA.
\end{IEEEbiography}

\begin{IEEEbiography}[{\includegraphics[width=1in,height=1.25in,clip,keepaspectratio]{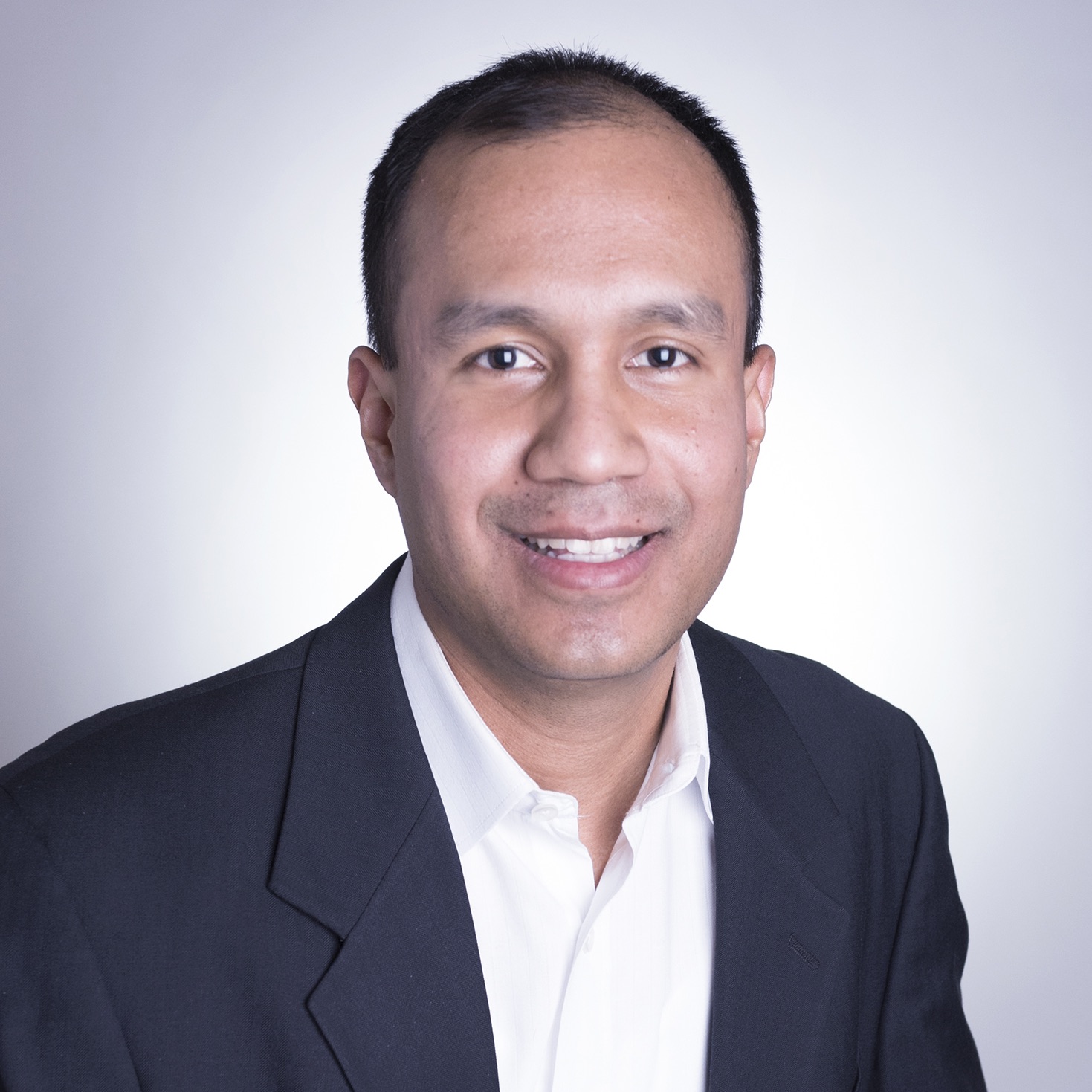}}]{Pablo Moriano}
(Senior Member, IEEE) received the B.S. and M.S. degrees in electrical engineering from Pontificia Universidad Javeriana, Colombia, and the M.S. and Ph.D. degrees in informatics from Indiana University Bloomington, Bloomington, IN, USA. He is a Staff Scientist with the Computer Science and Mathematics Division, Oak Ridge National Laboratory, Oak Ridge, TN, USA. His research interests include data science, network science, and cybersecurity, with a focus on the design of AI and statistical methods for anomaly detection in large-scale networked and cyber-physical systems. Applications of his work include intrusion detection, insider-threat analytics, and Internet-routing security. Dr. Moriano is also a Senior Member of the ACM and a member of SIAM.
\end{IEEEbiography}

\begin{IEEEbiography}
[{\includegraphics[width=1in,height=1.25in,clip,keepaspectratio]{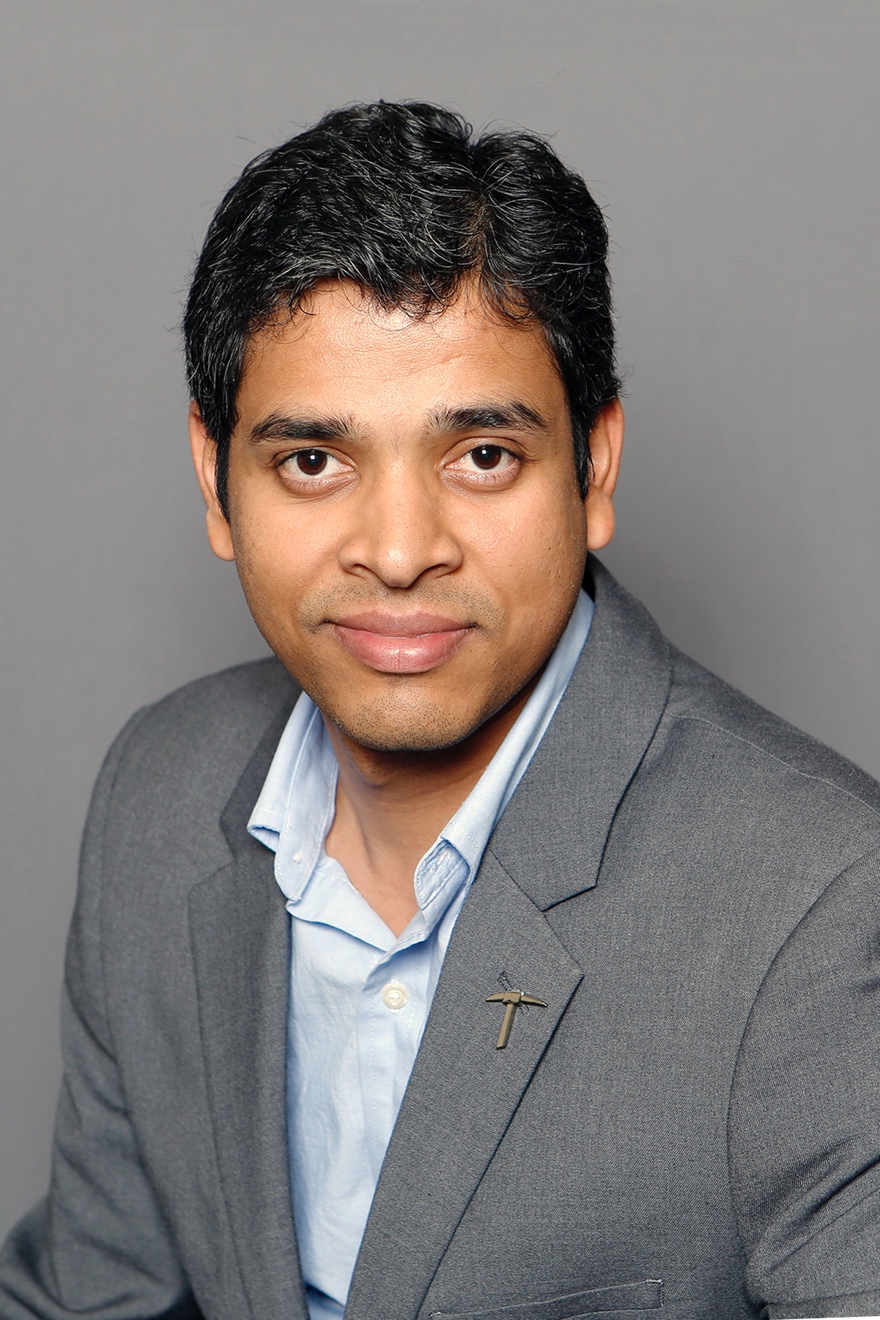}}]{Deepak Tosh}
(Senior Member, IEEE) is an associate professor of Computer Science at the University of Texas at El Paso. His research focuses on addressing various multi-disciplinary networking and cybersecurity challenges associated with critical national infrastructures, network virtualization, Industrial Internet of Things (IIoT), space cyber-physical systems, Blockchain, and tactical networking. His research has been supported by Department of Energy (DOE) and National Science Foundation (NSF) grants. He has authored/co-authored more than 90 peer-reviewed conference papers, book chapters, and journal papers. Two of his research works on Blockchain and data provenance received the “Top 50 Blockchain Papers in 2018” award at BlockchainConnect Conference, 2019. He is also a recipient of the prestigious NSF CAREER award, 2023.
\end{IEEEbiography}

\begin{IEEEbiography}
[{\includegraphics[width=1in,height=1.25in,clip,keepaspectratio]{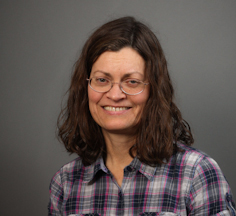}}]{Shirley V. Moore}
(Member, IEEE) received a B.A. degree in mathematics and chemistry from Indiana University, an M.Ed. degree from the University of Illinois, an M.S. in mathematics from Wichita State University, and M.S. and Ph.D. degrees in Computer Sciences from Purdue University. She is an Associate Professor in the Computer Science Department at the University of Texas at El Paso, USA. Her research is in parallel and high-performance computing, including distributed deep learning. Her focus is on performance optimization methodologies and tools. She is a Senior Member of ACM.
\end{IEEEbiography}

\vfill

\end{document}